\documentclass[a4paper,fleqn,usenatbib]{mnras}
\usepackage{newtxtext,newtxmath}
\usepackage[T1]{fontenc}
\usepackage{ae,aecompl}
\usepackage{graphicx}	% Including figure files
\usepackage{amsmath}	% Advanced maths commands
\usepackage{amssymb}	% Extra maths symbols
\usepackage{mathrsfs}
\usepackage{longtable}
\usepackage{hyperref}
\title[SN~2014C]{The changing-type SN~2014C may come from an 11-$M_\odot$ star stripped by binary interaction and violent eruption}

\author[N.-C. Sun et al.]{Ning-Chen Sun\thanks{E-mail: n.sun@sheffield.ac.uk}, 
Justyn R. Maund and Paul A. Crowther \\
Department of Physics and Astronomy, University of Sheffield, Hicks Building, Hounsfield Road, Sheffield S3 7RH, UK}

\date{Accepted XXX. Received YYY; in original form ZZZ}

\pubyear{2020}

\hypersetup{draft}

\begin{document}
\label{firstpage}
\pagerange{\pageref{firstpage}--\pageref{lastpage}}
\maketitle

%%%%%%%%
\begin{abstract}

SN~2014C was an unprecedented supernova (SN) that displayed a metamorphosis from Type~Ib to Type~IIn over $\sim$200~days. This transformation is consistent with a helium star having exploded in a cavity surrounded by a dense shell of the progenitor’s stripped hydrogen envelope. For at least 5 years post-explosion, the ejecta continued to interact with an outer, extended component of circumstellar medium (CSM) that was ejected even before the dense shell. It is still unclear, however, what kind of progenitor could have undergone such a complicated mass-loss history before it produced this peculiar SN. In this paper, we report a new analysis of SN~2014C's host star cluster based on data from the \textit{Hubble Space Telescope} (\textit{HST}). By carefully fitting its spectral energy distribution (SED), we derive a precise cluster age of 20.0$^{+3.5}_{-2.6}$~Myr, which corresponds to the progenitor's lifetime assuming coevolution. Combined with binary stellar evolution models, we find that SN~2014C's progenitor may have been an $\sim$11-$M_\odot$ star in a relatively wide binary system. The progenitor's envelope was partially stripped by \textit{Case~C} or \textit{Case~BC} mass transfer via binary interaction, followed by a violent eruption that ejected the last hydrogen layer before terminal explosion. Thus, SN~2014C, in common with SNe~2006jc and 2015G, may be a third example that violent eruptions, with mass-loss rates matching luminous blue variable (LBV) giant eruptions, can also occur in much lower-mass massive stars if their envelopes are partially or completely stripped in interacting binaries.

\end{abstract}

%%%%%%%%%
\begin{keywords}
supernovae: general -- supernovae: individual: 2014C -- stars: mass loss 
\end{keywords}

\defcitealias{M15}{M15}
\defcitealias{M17}{M17}
\defcitealias{T19}{T19}
\defcitealias{S20}{S20}

%%%%%%%%%%%
\section{Introduction}
\label{intro.sec}

Massive stars experience a wide variety of mass loss during their lifetime, which can significantly affect their evolution and end fates \citep{Smith2014araa}. Owing to their intense radiation, hot massive stars have much stronger line-driven winds than lower-mass stars \citep{Bestenlehner2014, Bestenlehner2020}. The stellar winds may become even stronger as they evolve into the red supergiant (RSG) stage, especially when their envelopes become dynamically unstable and develop large-amplitude pulsations \citep[e.g.][]{Yoon2010}. A subset of stars with initial masses $M_{\rm ini}$~$\ge$~25~$M_\odot$ may undergo a special phase as luminous blue variables (LBVs) with a high degree of mass loss \citep{Vink2012}. LBVs are very unstable and exhibit irregular variabilities, the most pronounced of which is the so-called giant eruptions. During giant eruptions, LBVs increase their bolometric luminosities for months to years and suffer from extreme mass loss with mass-loss rates of $\dot{M}$~$\gtrsim$~10$^{-2}$~$M_\odot$~yr$^{-1}$ (e.g. the Great Eruption of $\eta$~Carina and P Cygni's 1600 AD eruption; \citealt{Humphreys1999, Smith2006, Smith2011}). LBVs were believed to be a transitional phase from the main-sequence (MS) to the Wolf-Rayet (WR) stage \citep{Conti1976, Massey2003}. Accumulating evidence, however, shows that LBVs may have much more heterogenous origins than previously thought \citep[e.g.][]{Smith2019}. Despite their importance, the diverse mass-loss processes are still a major uncertainty in understanding the evolution of massive stars.

SNe provide a unique opportunity to probe the mass loss of their progenitors at the latest evolutionary stages before core collapse. Many massive stars exhibit enhanced mass loss before their final explosion. For a subset of them, the pre-SN mass loss is so intense that the progenitors are surrounded by very dense CSM; after SN explosion, narrow/intermediate-width (10$^2$--10$^3$~km~s$^{-1}$) emission lines can be produced as the SN ejecta catch up and collide with the dense CSM \citep{Chugai2004}. SNe with such spectral features are classified as Type~IIn (CSM is hydrogen-rich) or Type~Ibn (CSM is hydrogen-poor and helium-rich; \citealt{Pastorello2008}). Type~IIn seems to be more heterogeneous than Type~Ibn \citep{Hosseinzadeh2017}, and some Type~IIn SNe may arise from massive RSGs with superwinds \citep[e.g. SNe~1998Z and 2005ip;][]{Smith2017sn2005ip}. For many other Type~IIn SNe, however, pre-SN mass-loss rates and CSM expansion velocities are consistent with those of LBV giant eruptions, suggesting that their progenitors may be massive LBVs \citep{Smith2017book}. While this is quite surprising, progenitor detections for a handful of Type~IIn SNe confirms that LBVs can indeed directly undergo core collapse (e.g. SN~2005gl, \citealt{Gal-Yam2007}, \citealt{Gal-Yam2009}; SN~2009ip, \citealt{Mauerhan2013}). Coupled stellar evolution and atmospheric modelling also shows that some stars may exhibit LBV-like luminosity, spectra and chemical composition just before their final SN explosion \citep{Groh2013a, Groh2013b}.

For Type~Ibn SNe, the pre-SN mass loss is also comparable to LBV giant eruptions in terms of their mass-loss rates and CSM expansion velocities \citep{Smith2017book}. For the class prototype SN~2006jc, a pre-SN outburst was even directly detected as an optical transient, with peak brightness and duration similar to those of LBV giant eruptions \citep{Pastorello2007}. However, LBVs are still rich in hydrogen but Type~Ibn SN progenitors are not. Early studies generally assumed that Type~Ibn SNe arise from massive WR stars, which still have some residual instability after they descend from the LBV phase \citep[e.g.][]{Foley2007, Pastorello2008}. On the other hand, the recent work of \citealt{Shivvers2017} and \citealt{S20} (hereafter \citetalias{S20}) rule out massive WR progenitors for the Type~Ibn SNe~2006jc and 2015G and argue for much lower-mass progenitors stripped via binary interaction. This led \citetalias{S20} to conclude that violent pre-SN mass loss that resembles LBV giant eruptions can also occur in lower-mass massive stars (8~$<$~$M_{\rm ini}$~$<$~25~$M_\odot$) if their envelopes are removed in interacting binary systems. Such a conclusion may confirm the theoretical prediction that envelope removal can aid the occurrence of pre-SN eruption since, without a massive envelope, the core convection-excited waves are easier to reach and trigger an outburst at the stellar surface \citep{Fuller2017, Fuller2018}.

SN~2014C is an unprecedented SN that exhibited a transformation from Type~Ib to Type~IIn (\citealt{M15}, hereafter \citetalias{M15}). It had a very typical Type~Ib spectrum soon after its explosion, suggesting that the progenitor was a stripped helium star with little remaining hydrogen. Over a timescale of 200~days, however, its spectrum started to show prominent H$\alpha$ emission with an intermediate width of 1000-2000~km~s$^{-1}$. This indicates that the SN ejecta was interacting with a dense CSM shell, which was, however, rich in hydrogen. The interaction has also produced strong radio \citep{A17, B18} and X-ray emission (\citealt{M17}, hereafter \citetalias{M17}). Detailed analysis shows that the shell is located at $\sim$6~$\times$~10$^{16}$~cm from the progenitor and has a thickness of 10$^{16}$~cm, a density of 2~$\times$~10$^6$~cm$^{-3}$ and a mass of 1.0--1.5~$M_\odot$ \citepalias{M17}. The mass ejection that produced this shell had a mass-loss rate of (3--5)~$\times$~10$^{-2}$~$M_\odot$~yr$^{-1}$, assuming an ejection velocity of 100~km~s$^{-1}$ \citepalias{M15}. These characteristics are very similar to those of LBV giant eruptions. Thus, SN~2014C serves as another important laboratory to study the violent eruptions of stripped stars before their core collapse.

The dense shell is not the only CSM component around SN~2014C's progenitor. \citet[][hereafter \citetalias{T19}]{T19} carried out a long-term monitoring campaign of SN~2014C in the near- to mid-infrared wavelengths. They find that ejecta-CSM interaction was still on-going even at 5 years post-explosion. This suggests that there was an extended CSM component outside the dense shell out to at least $\sim$2~$\times$~10$^{17}$~cm from the progenitor. This component has a different density profile from that of the dense shell and corresponds to a lower (but still very high) mass-loss rate of (1.1--2.6)~$\times$~10$^{-3}$~$M_\odot$~yr$^{-1}$. Thus, SN~2014C's progenitor has experienced a very complicated mass-loss history. However, it is still unclear what kind of progenitor could form such a complex CSM configuration and produce SN~2014C with such a peculiar transformation from Type~Ib to Type~IIn.

\citetalias{M15} found SN~2014C to be spatially coincident with a star cluster, which very likely hosts the SN progenitor. Note that, perhaps quite surprisingly, core-collapse SNe rarely explode in bright star clusters \citep{Smartt2009}; some of the progenitors may have been ejected from their host star clusters while some others may not be born in star clusters at all (recent studies show that massive stars form not only in compact clusters but also in much looser groups that are hierarchically structured; e.g. \citealt{Sun2017a, Sun2017b, Sun2018}). Thus, the host star cluster of SN~2014C (which we shall refer to as \textit{Cluster~A}) provides a precious opportunity to infer the properties and the pre-SN evolution of the SN progenitor. In particular, the age of \textit{Cluster~A} should correspond to the lifetime of the progenitor star, assuming they are coeval with each other. Following this idea, \citetalias{M15} derived an age of 30--300~Myr based on pre-explosion observations in the \textit{B}, \textit{V} and \textit{R} bands with the ground-based \textit{Subaru} telescope. \citetalias{M15} prefer a value at the lower range of potential ages given the strong H$\alpha$ emission detected in the \textit{F658N} band by \textit{HST}.

In this paper, we report a new analysis of SN~2014C's host star cluster. We improve the age estimate in three aspects. First of all, we compile an \textit{HST} dataset of SN~2014C over a much wider wavelength range from the UV to the near-infrared. The UV filters are very sensitive to the cluster age and thus can put much tighter constraints. Secondly, \textit{Cluster~A}, in the lower-resolution images, may be confused by the nearby sources in its close vicinity. The new data, with a much higher spatial resolution, allow us to resolve and remove the contamination of the nearby sources. Thirdly, we carry out a more detailed modelling of the cluster's SED from not only its the stellar population but also its gaseous nebula. The emission lines from the nebula can affect the brightness in the broad bands (e.g. H$\alpha$ in the \textit{R} band) and can lead to systematic errors if they are not accounted for. On the other hand, the nebular lines serve as sensitive age indicators since they are powered by the ionising radiation from the massive stars. Our aim is to derive a precise estimate for \textit{Cluster~A}'s age, and in turn, to infer the property and the pre-SN evolution of SN~2014C's progenitor.

Throughout this paper, we assume a distance of 14.7~$\pm$~0.6~Mpc \citep{dis.ref} and a solar metallicity for SN~2014C, both consistent with \citetalias{M15}. This paper is structured as follows: Section~\ref{data.sec} is a summary of the data used in this work. Section~\ref{phot.sec} describes how we measure the brightness of the host star cluster and in Section~\ref{property.sec} we try to infer its properties. Section~\ref{progenitor.sec} further explores the properties and the pre-SN evolution of SN~2014C's progenitor. We finally close this paper with a summary and discussion. 

%%%%%%%%%%%%
\section{Observations}
\label{data.sec}

\begin{table*}
\label{obs.tab}
\center
\caption{\textit{HST} Observations of SN~2014C}
\begin{tabular}{cccccc}
\hline
\hline
Program & Date & Phase$^{\rm f}$ & Instrument & Filter & Exposure\\
ID & (UT) & (yr) &   &    & Time (s) \\
\hline
11966$^{\rm a}$ 
& 2009-01-01.2 & -5.0 & \textit{     WFPC2} & \textit{F658N} &  600.0 \\
& 2009-01-01.2 & -5.0 & \textit{     WFPC2} & \textit{F658N} &  600.0 \\
& 2009-01-01.2 & -5.0 & \textit{     WFPC2} & \textit{F658N} &  600.0 \\
\hline
14202$^{\rm b}$
& 2015-08-22.4 &  1.6 & \textit{WFC3/UVIS2} & \textit{F438W} & 1380.0 \\
& 2015-08-22.4 &  1.6 & \textit{WFC3/UVIS2} & \textit{F814W} & 1350.0 \\
& 2015-08-22.5 &  1.6 & \textit{WFC3/UVIS2} & \textit{F275W} & 1944.0 \\
& 2015-08-22.5 &  1.6 & \textit{WFC3/UVIS2} & \textit{F657N} & 1920.0 \\
& 2015-08-22.6 &  1.6 & \textit{WFC3/UVIS2} & \textit{F657N} &  480.0 \\
\hline
14668$^{\rm c}$
& 2016-10-12.1 &  2.7 & \textit{ WFC3/UVIS} & \textit{F336W} &  780.0 \\
& 2016-10-12.1 &  2.7 & \textit{ WFC3/UVIS} & \textit{F555W} &  710.0 \\
\hline
14762$^{\rm d}$
& 2017-01-13.3 &  3.0 & \textit{WFC3/UVIS2} & \textit{F300X} & 1200.0 \\
& 2017-01-13.3 &  3.0 & \textit{WFC3/UVIS2} & \textit{F475X} &  350.0 \\
\hline
15166$^{\rm c}$
& 2018-01-14.5 &  4.0 & \textit{ WFC3/UVIS} & \textit{F336W} &  780.0 \\
& 2018-01-14.5 &  4.0 & \textit{ WFC3/UVIS} & \textit{F555W} &  710.0 \\
\hline
15645$^{\rm e}$
& 2019-05-02.8 &  5.3 & \textit{   ACS/WFC} & \textit{F814W} & 2152.0 \\
\hline
\multicolumn{6}{l}{PIs: (a) M. Reganl; (b) D. Milisavljevic; (c) A. Filippenko; (d) J. R. Maund; (e) D. Sand.} \\
\multicolumn{6}{l}{(f): Phase is with respect to \textit{V}-band maximum on 2014-01-13.} \\
\end{tabular}
\end{table*}

We used a series of \textit{HST} observations of SN~2014C, a complete list of which is provided in Table~1. They were conducted by the \textit{Wide Field and Planetary Camera 2} (\textit{WFPC2}), \textit{Wide Field Camera 3} (\textit{WFC3}) \textit{Ultraviolet-Visible} (\textit{UVIS}) channel, and the \textit{Advanced Camera for Surveys} (\textit{ACS}) \textit{Wide Field Channel} (\textit{WFC}). The observations span a long period from five years before to more than five years post-explosion of SN~2014C and cover a long wavelength range from the UV to the \textit{F814W} band. They also include observations in two narrow bands: the \textit{WFPC2/F658N} band (Program~11966), which covers the H$\alpha$~$\lambda$6563 and the [N~{\small II}]~$\lambda$6584 lines\footnote{Centred on 6591~\AA\ and with a width of 28.5~\AA, the \textit{WFPC2/F658N} filter can only cover the [N~{\small II}]~$\lambda$6584 line in the rest frame. Due to the recession velocity of \textit{Cluster~A} \citepalias[990~km~s$^{-1}$;][]{M15}, however, this filter can also cover the H$\alpha$~$\lambda$6563 line in the redshifted spectrum.}, and the \textit{WFC3/F657N} band (Program~14202), which covers the H$\alpha$~$\lambda$6563 and the [N~{\small II}]~$\lambda\lambda$6548, 6584 lines.

For the \textit{WFPC2} observations (Program~11966), combined and well calibrated images were retrieved from the Hubble Legacy Archive\footnote{\url{https://hla.stsci.edu}} and used in this work without any further reduction. For the other observations, we retrieved the data from Mikulski Archive for Space Telescope\footnote{\url{https://archive.stsci.edu}} and manually combined the dithered exposures for each filter with the the ASTRODRIZZLE package\footnote{\url{ http://drizzlepac.stsci.edu}}. In doing this, we set \texttt{driz\_cr\_grow = 3} for more efficient removal of cosmic rays; all other drizzle parameters were kept unchanged as in the standard calibration pipeline \citep{acs.ref, wfc3.ref}.

%%%%%%%%%%%
\section{Photometry of the Host Star Cluster}
\label{phot.sec}

\begin{figure*}
\centering
\includegraphics[scale=0.45, angle=0]{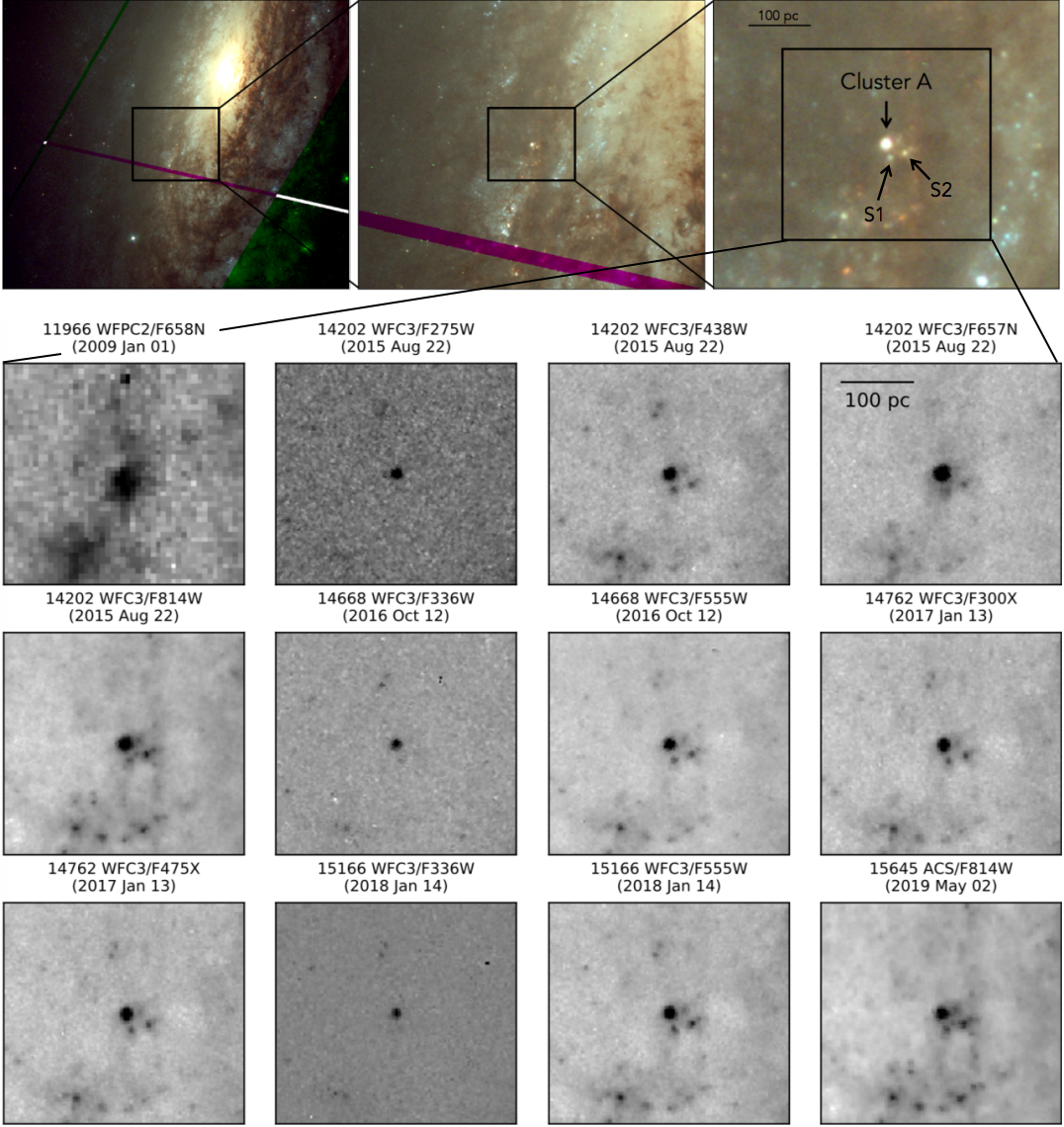}
\caption{\textit{HST} observations of the site of SN~2014C. The top panels are three-colour composite images of the \textit{F438W} (blue), \textit{F555W} (green) and \textit{F814W} (red) bands; the \textit{F438W} and \textit{F814W} images are from Program~14202 and the \textit{F555W} image is from Program~14668; the greenish and reddish regions are not covered by all three images. The other panels show the SN site from individual \textit{HST} observations; the images are centred on \textit{Cluster~A} (i.e. SN~2014C's host star cluster) and have been set to the same scale (the bar in the top-right panel corresponds to 100~pc). All panels are aligned such that North is up and East is to the left.}
\label{image.fig}
\end{figure*}

\begin{figure*}
\centering
\includegraphics[scale=0.8, angle=0]{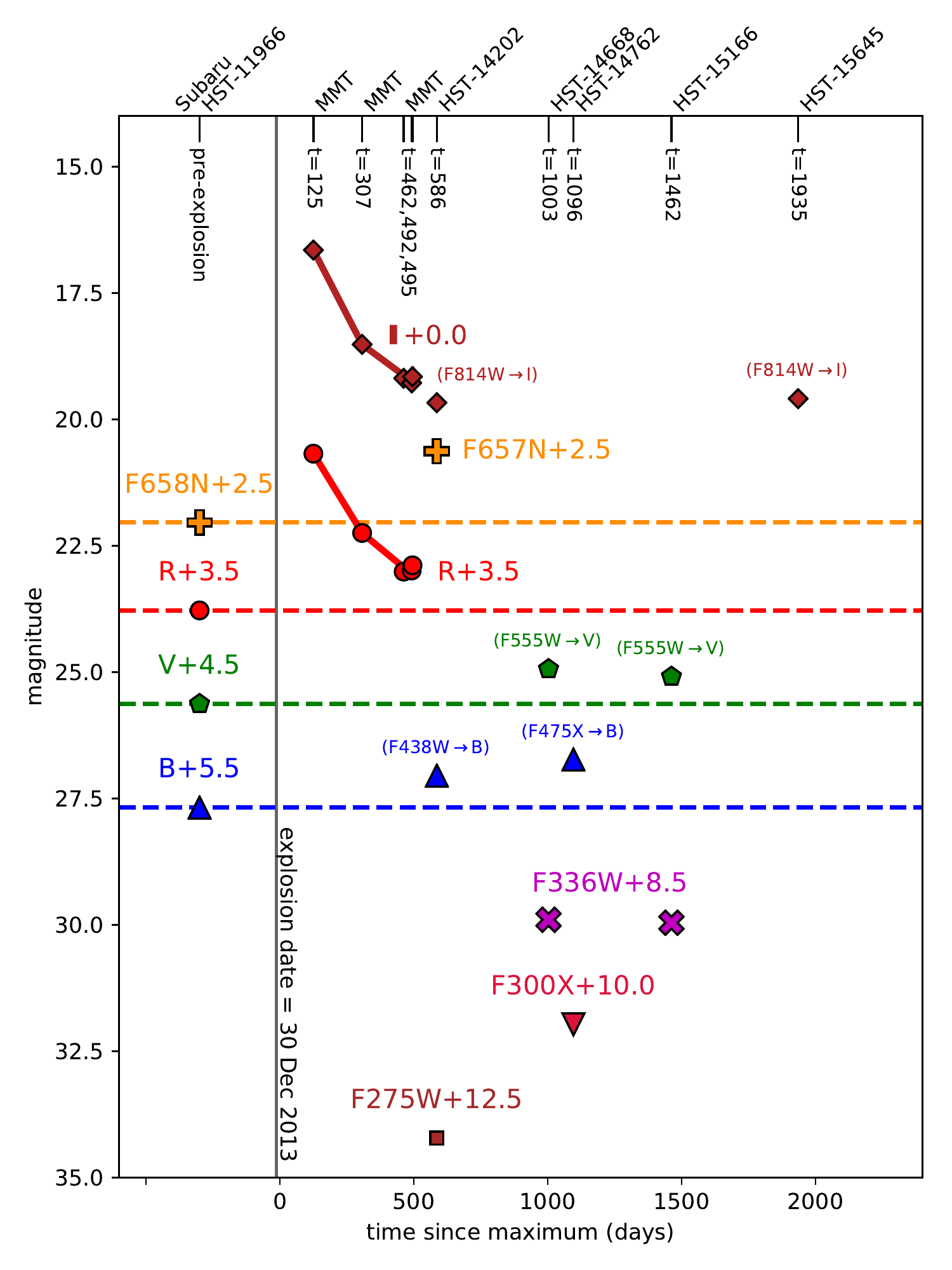}
\caption{Magnitudes of \textit{Cluster~A} in different bands and at different epochs, when the light of SN~2014C may or may not be significant. Data points connected by solid lines are the late-time light curves of SN~2014C from \citetalias{M17}. The horizontal dashed lines indicate the possible magnitudes of the host star cluster itself, i.e. without significant light contamination from SN~2014C.}
\label{curve.fig}
\end{figure*}

\begin{table*}
\label{mag.tab}
\center
\caption{Magnitudes of \textit{Cluster~A} from GALFIT results}
\begin{tabular}{cccccccccc}
\hline
\hline
Phase$^{\rm a}$ & {\it F275W} & {\it F300X} & {\it F336W} & {\it F438W} & {\it F475X} & {\it F555W} & {\it F657N} & {\it F658N} & {\it F814W} \\
(yr) & (mag) & (mag) & (mag) & (mag) & (mag) & (mag) & (mag) & (mag) & (mag) \\
\hline
$-$5.0 & ...  & ...  & ...  & ...  & ...  & ...  & ...  & 19.54 (0.09) & ...  \\
1.6 & 21.72 (0.01) & ...  & ...  & 21.58 (0.01) & ...  & ...  & 18.13 (0.01) & ...  & 19.67 (0.08) \\
2.7 & ...  & ...  & 21.40 (0.03) & ...  & ...  & 20.51 (0.02) & ...  & ...  & ...  \\
3.0 & ...  & 21.97 (0.01) & ...  & ...  & 20.76 (0.03) & ...  & ...  & ...  & ...  \\
4.0 & ...  & ...  & 21.46 (0.04) & ...  & ...  & 20.68 (0.02) & ...  & ...  & ...  \\
5.3 & ...  & ...  & ...  & ...  & ...  & ...  & ...  & ...  & 19.59 (0.07) \\
\hline
\multicolumn{10}{l}{(a): Phase is with respect to \textit{V}-band maximum on 2014-01-13.} \\
\end{tabular}
\end{table*}

Figure~\ref{image.fig} shows the site of SN~2014C as imaged by \textit{HST} in different filters and at different epochs. \textit{Cluster~A} is clearly revealed in all the images, and we derived its photometry by modelling all the sources in a 1.6\arcsec~$\times$~1.6\arcsec box region centred on it with the GALFIT package \citep{galfit1.ref, galfit2.ref}. The details can be found in Appendix~\ref{galfit.sec}. Note that, to the south and southwest of \textit{Cluster~A}, there are two additional sources (\textit{S1} and \textit{S2}, respectively) that are fainter but still visible in most of the images. \textit{S1} and \textit{S2} are associated with significant \textit{WFC3/F657N} emission; in the \textit{WFPC2/F658N} image, however, they are confused with and cannot be distinguished from \textit{Cluster~A} due to the poor spatial resolution (SN~2014C falls on the \textit{WF3} chip of \textit{WFPC2}, which has a pixel scale of 0.1$\arcsec$). In Appendix~\ref{line.sec}, we have estimated the contributions of \textit{S1} and \textit{S2} in the \textit{WFPC2/F658N} band, which are comparable to the brightness of \textit{Cluster~A}. They are then removed from the measured brightness. The derived and decontaminated magnitudes (Vega magnitude system is used throughout this paper) of \textit{Cluster~A} are listed in Table~2 for all the \textit{HST} observations.

The host galaxy of SN~2014C was also observed by the 8.2m \textit{Subaru} Telescope with the \textit{Suprime-Cam} instrument \citep{subaru.ref} before the explosion of SN~2014C. \textit{Cluster~A} is apparent in the \textit{Subaru} images, and \citetalias{M15} derived its magnitudes with PSF photometry as $m_B$~=~22.18~$\pm$~0.13, $m_V$~=~21.13~$\pm$~0.09, and $m_R$~=~20.28~$\pm$~0.06. We have independently performed data reduction and photometry on the \textit{Subaru} images. Our results are consistent with those of \citetalias{M15} within errors. Strictly speaking, \textit{Cluster~A} is also confused with \textit{S1}, \textit{S2} and other nearby sources on the \textit{Subaru} images. However, they are much fainter than \textit{Cluster~A} in the broad optical bands. Moreover, PSF photometry assigns lower weights to pixels that are far from the PSF centre. Thus, we do not perform any decontamination in the \textit{B}-, \textit{V}- and \textit{R}-band photometry.

\subsection{Was the SN still bright?}

Apart from the \textit{Subaru} observations and \textit{HST} Program~11966, all other observations were acquired after the explosion of SN~2014C. Thus, it is very important to check whether the radiation from the SN itself was still significant. To this end, we compare the magnitudes of \textit{Cluster~A} at different epochs (Fig.~\ref{curve.fig}), which may or may not be contaminated by SN~2014C. To aid the comparison, the \textit{HST} native-filter magnitudes in the optical broad bands have been converted to the standard Johnson-Cousins system (\textit{F438W} and \textit{F475X} to the \textit{B} band, \textit{F555W} to the \textit{V} band, and \textit{F814W} to the \textit{I} band) based on transformation relations reported by \citet{Sirianni2005} and \citet{Harris2018}. The UV-band and narrow-band magnitudes are kept unchanged as in their native filters. In Fig.~\ref{curve.fig}, we also show the late-time light curves of SN~2014C as observed by the 6.5m \textit{MMT} telescope \citepalias{M17}, which have been transformed from the SDSS photometric system to the Johnson-Cousins system based on the relation of \citet{Jordi2006}. Note, however, that the transformation relations used here are calibrated based on normal stars; the presence of strong emission lines may lead to systematic uncertainties.

In the \textit{B} and \textit{V} bands, the late-time brightness was brighter than the pre-explosion level from $t$~=~586 to 1462~days. This suggests that the radiation from SN~2014C was still significant in these bands. As SN~2014C experienced strong ejecta-CSM interaction, its late-time spectra exhibit prominent emission lines, the strongest of which include H$\alpha$~$\lambda$6563, H$\beta$~$\lambda$4861, [O {\small III}]~$\lambda$4363 and [O {\small III}]~$\lambda\lambda$4959,~5007, etc. (\citetalias{M15}; \citealt{A17}). The \textit{B}-band magnitude converted from \textit{F475X} at $t$~=~1096~days is slightly brighter than that converted from \textit{F438W} at $t$~=~586~days. However, this does not necessarily suggest an increase in the SN's \textit{B}-band brightness; this could also be due to the strong emission lines around $\sim$5000~\AA, which are covered by the extremely wide \textit{F475X} band but not by the \textit{F438W} band. The \textit{V}-band (or the \textit{F555W}-band) magnitude faded by 0.17~mag from $t$~=~1003 to 1462~days, suggesting that the brightness of SN~2014C was on a decline. Thus, the late-time \textit{HST} observations in the \textit{F438W}, \textit{F475X} and \textit{F555W} bands are still subject to significant emission from the SN and cannot be directly used to infer the properties of the host star cluster.

In contrast, the \textit{I}-band light curve remains almost constant from $t$~=~586 to 1935~days (which corresponds to a time span of 3.7~years). The magnitude difference is only 0.08~mag, which is within the combined photometric error (0.11~mag). We also find that the \textit{I}-band magnitude at $t$~=~586~days is consistent with the extrapolation of SN~2014C's earlier \textit{MMT} light curve. Thus, it is possible that SN~2014C has faded significantly in the \textit{I} band by $t$~=~586~days, after which the light from \textit{Cluster~A} dominates the \textit{I}-band brightness. We also note that the emission lines in the \textit{I} band are much weaker than those in the \textit{B}, \textit{V} and \textit{R} bands (\citetalias{M15}; \citealt{A17}). To be conservative, however, we do not exclude the possibility that the SN was still bright in the \textit{I} band and we regard the measured \textit{F814W} brightness as an upper limit for \textit{Cluster~A}.

Similarly, the \textit{F336W}-band magnitude also remains almost constant from $t$~=~1003 to 1462~days (which corresponds to a time span of 1.3~years). The magnitude difference is only 0.06~mag, which is comparable to the combined photometric error (0.05~mag). Note that during the same period the \textit{F555W}-band magnitudes faded by 0.17~mag. Yet, SN~2014C, like many other interacting SNe, exhibits a blue pseudo-continuum, which may arise from blended iron lines produced by ejecta-CSM interaction (\citetalias{M15}; see also, e.g., SN~2006jc, \citealt{Pastorello2007}; \citealt{Bufano2009}). As a result, we cannot rule out the possibility that the \textit{F336W} band is contaminated by the pseudo-continuum emission from the SN. Thus, we regard the \textit{F336W}-band brightness also as an upper limit for \textit{Cluster~A}.

The \textit{F275W} and \textit{F300X} bands were only observed at a single epoch. Both bands cover the [Mg~{\small II}] doublets at $\sim$2800~\AA; the \textit{F300X} band also covers the [Fe~{\small II}] line at $\sim$2950~\AA. These two lines are important coolants and often appear to be very strong in emission for interacting SNe \citep{Fesen1999, Fransson2005, Bufano2009}. Thus, \textit{F275W} and \textit{F300X} bands may be contaminated by the SN light, although it is difficult to estimate the amount of its contribution. Still, they place important upper limits for the brightness of \textit{Cluster~A} at UV wavelengths.

In summary, SN~2014C's light was still significant until at least 4 years post explosion (the last epoch of \textit{F555W} observation). This is consistent with the conclusion of \citetalias{T19} that SN~2014C experiences on-going interaction between its ejecta and CSM at large distances. Still, the \textit{HST} observations provide very important constraints on \textit{Cluster~A}'s SED from the UV to the \textit{F814W} band, a much wider wavelength coverage than the pre-explosion observations in the \textit{B}, \textit{V}, \textit{R} and \textit{F658N} bands. Thanks to the higher spatial resolution provided by the new data, the measured SED of \textit{Cluster~A} is now much cleaner, i.e., without significant contamination from its nearby sources.

%%%%%%%%%%%%%%%%%%%%%%
\section{Properties of the Host Star Cluster}
\label{property.sec}

\begin{figure*}
\centering
\includegraphics[scale=0.5, angle=0]{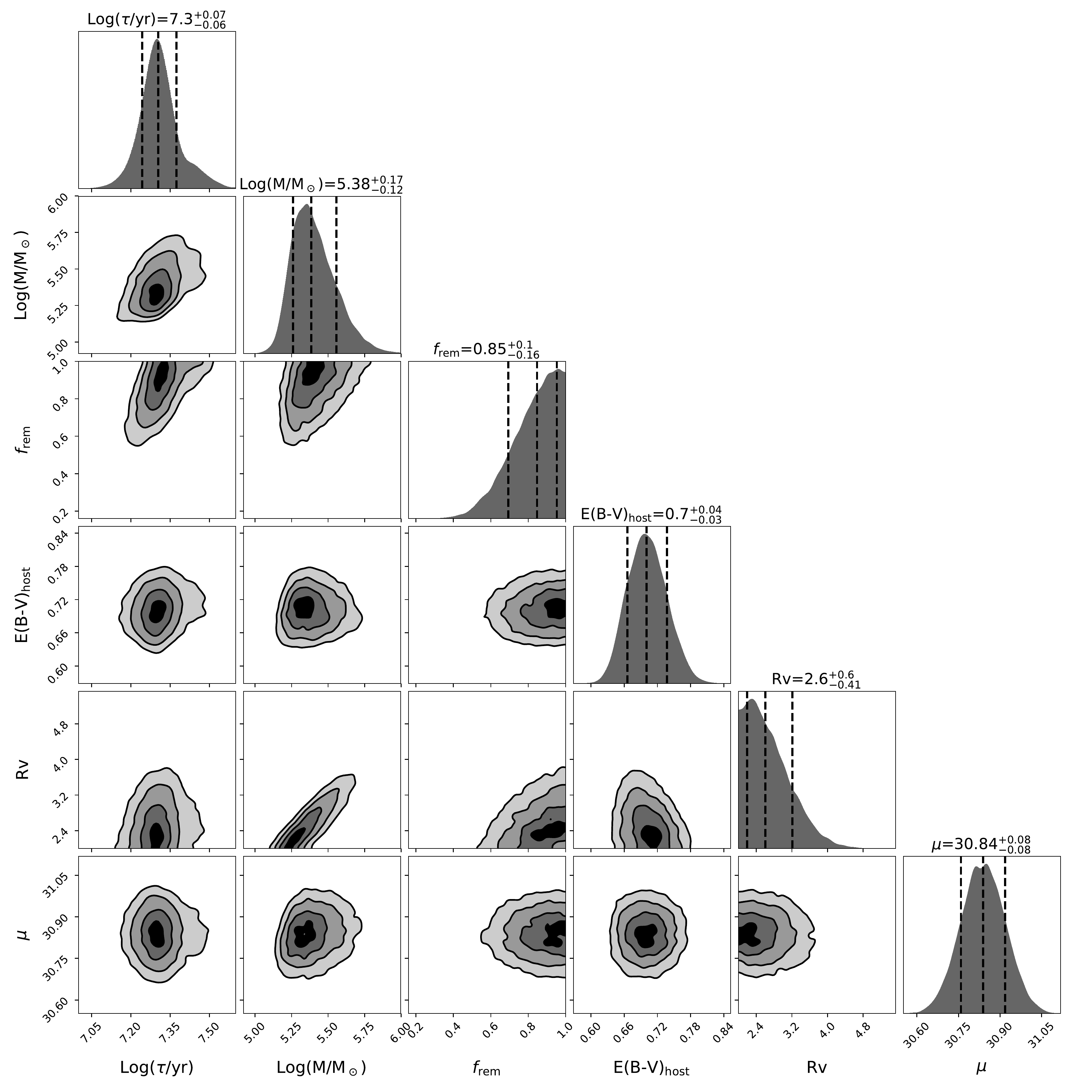}
\caption{Marginalised posterior probability distributions over one (with median values and 1-$\sigma$ credible intervals) and two dimensions (with 0.5, 1.0, 1.5 and 2.0-$\sigma$ contours from inside to outside).}
\label{post1.fig}
\end{figure*}

\begin{figure}
\centering
\includegraphics[scale=0.455, angle=0]{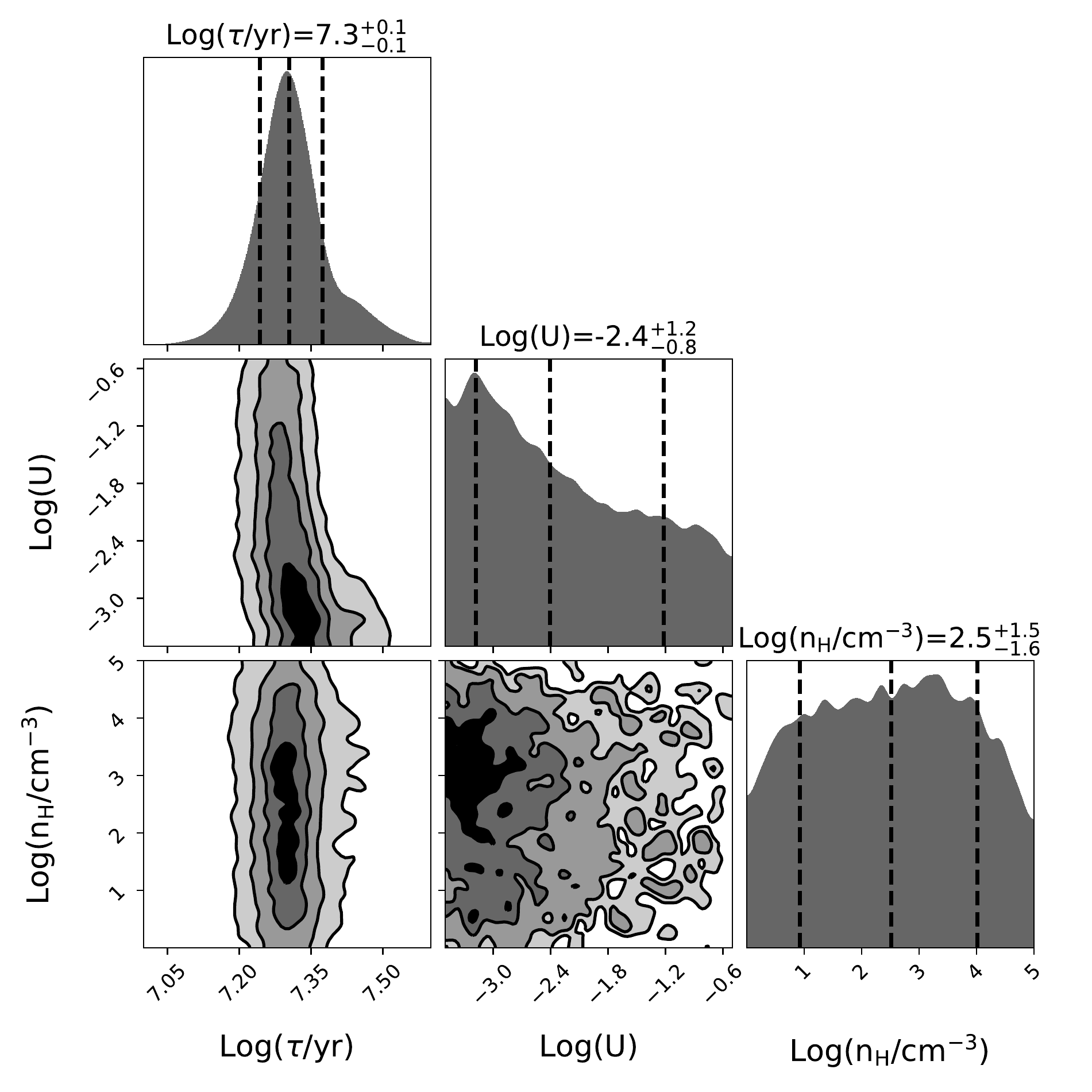}
\caption{Same as Fig.~\ref{post1.fig} but for parameters of cluster age, ionisation parameter and hydrogen density.}
\label{post2.fig}
\end{figure}

\begin{figure*}
\centering
\includegraphics[scale=0.7, angle=0]{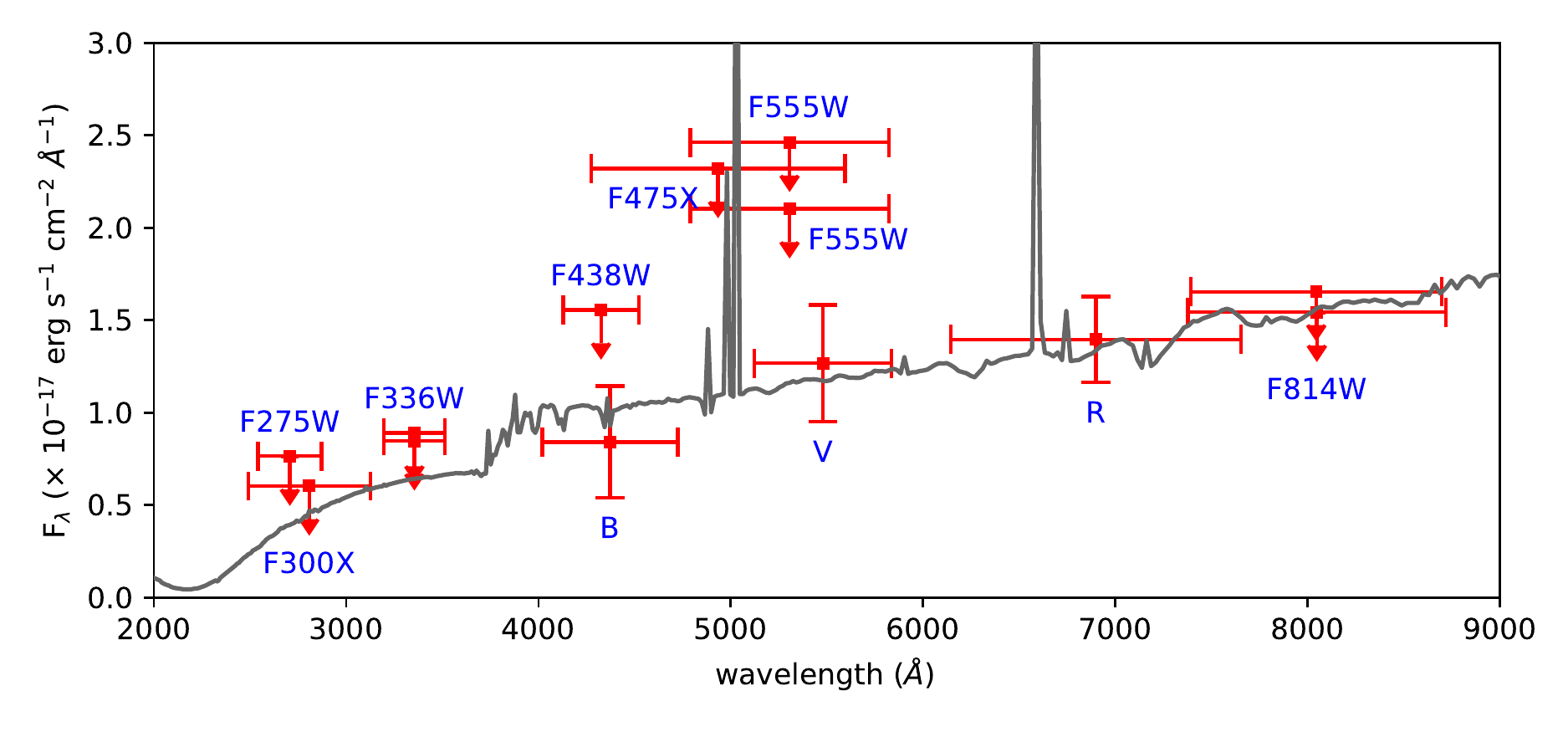}
\caption{The observed SED (red data points) and best-fitting model spectrum (black line) of \textit{Cluster~A}. The data points are centred on the Pivot wavelength of each band, and the horizontal error bars represent the root-mean-square band widths of the filters. The vertical error bars reflect the $\pm$3$\sigma$ photometric errors. Data points for the narrow bands are not displayed but they agree with the best-fitting model spectrum.}
\label{spec.fig}
\end{figure*}

We further try to infer the properties of \textit{Cluster~A} by fitting model spectra to its observed SED. Note that \textit{Cluster~A} is still associated with a considerable amount of (ionised) gas, given its bright emission in the \textit{F658N} band. The bright nebular line emission will significantly influence its broad-band SED (e.g. H$\alpha$ in the \textit{R} band) and must be modelled properly. To do this, we make use of the BPASS binary stellar population models \citep[version~2.1;][]{bpass.ref} to simulate the stellar component and the CLOUDY photoionisation code \citep[version~17.01;][]{cloudy.ref} to calculate the strength of nebular lines.

BPASS provides a large set of binary evolution models and spectral synthesis for stellar populations of various ages, metallicities and stellar initial mass functions. Binary evolution is calculated on a grid of initial mass ratio $M_2$/$M_1$~=~0.1--0.9 in steps of 0.1, orbital period log($P$/days)~=~0.0--4.0 in steps of 0.2, and 68 values of primary stellar mass from 0.1 to 300~$M_\odot$. While some binaries never interact during their lifetimes, some binaries do experience Roche-lobe overflow (RLOF) or Common-Envelope (CE) evolution (the numerical recipe is described in \citealt{bin.ref}). Each population contains binaries with a flat distribution in the initial mass ratio and log-period. To accurately model \textit{Cluster~A}'s SED, it is very important to consider the effect of interacting binaries. Binary interaction can strip a star's envelope and prevent it from evolving into a cool RSG. As a result, the stripped star remains very hot and continues to provide ionising photons that power the nebular emission of the surrounding gas. As shown in a number of studies \citep[e.g.][]{Xiao2018, Xiao2019}, using population models without interacting binaries can significantly affect the estimate of physical properties.

Thus, we use the binary-star population models of BPASS to model the stellar component of \textit{Cluster~A} while assuming a solar metallicity and a \citet{Kroupa2001} initial mass function with a maximum stellar mass of 300~$M_\odot$. Synthetic SEDs are then retrieved (from the file \texttt{spectra-bin-imf135\_300.z020.dat.gz}) for stellar populations (which have been normalised to a total mass of 10$^6$~$M_\odot$) as a function of age from log($\tau$/yr)~=~6.0 to 11.0 in steps of 0.1.

We then use CLOUDY to model the nebular emission with the BPASS SEDs as the ionising source. For simplicity, we firstly assume that the gas nebula is a spherical shell around the stars and that the nebula is dust-free, ionisation-bound, and has a uniform, non-evolving density. We match the nebular metallicity with that of the stellar component and set the hydrogen density to be log($n_{\rm H}$/cm$^{-3}$)~=~0.0 to 5.0 in steps of 1.0. This density range is consistent with the observed electron density for H~{\small II} regions \citep[e.g.][]{Hunt2009}. The intensity of the ionising flux is characterised by the ionisation parameter:
\begin{equation}
U = \dfrac{Q(H)}{4 \pi r_{0}^{2} n_{\rm H} c}
\end{equation}
where $Q(H)$ is the hydrogen-ionising photon luminosity of the source, $r_0$ the inner radius of the nebula, and $c$ is the speed of light. This parameter is the dimensionless ratio of hydrogen-ionising photon to hydrogen densities. We set log($U$)~=~$-$3.5 to $-$0.5 in steps of 0.3, which matches the observed range for H~{\small II} regions \citep[e.g.][]{Rigby2004}. For each combination of age, hydrogen density and ionisation parameter, we run the CLOUDY code to simulate the photoionisation process in the gas nebula. As an output, we obtain the synthetic spectrum for each model, which includes not only the emission from the stellar component but also that from the nebular component.

In practice, the ionising photons from massive stars may not all have a chance to interact with the surrounding gas. For example, the gas may not cover the full sky as viewed from the central stars. Alternatively, the gas shell may be optically thin and some ionising photons can escape without ionising any atoms. Thus, we use another parameter, $f_{\rm rem}$, the fraction of remaining ionising photons that have not escaped, to characterise this ``photon leakage effect" \citep{Xiao2018, Xiao2019}. The value of $f_{\rm rem}$ can vary between 0 and 1, and the line fluxes are scaled linearly with $f_{\rm rem}$ to a good approximation.

We apply a Galactic reddening of $E(B-V)_{\rm mw}$~=~0.08~mag \citep{sfd.ref} with a Galactic extinction law \citep[$R_V$~=~3.1;][]{Fitzpatrick2004}. SN~2014C occurred in the dusty disk of its host galaxy (Fig.~\ref{image.fig}) and the interstellar reddening from its host galaxy is not negligible. Thus, we further redden the synthetic spectra by reddening values of $E(B-V)_{\rm host}$~=~0.6 to 1.0 in steps of 0.05. We assume a \citet{Fitzpatrick2004} extinction law, but leave the total-to-selective extinction ratio $R_V$~=~$A_V$/$E(B-V)$ as a free parameter, ranging from 2.0 to 6.0 in steps of 0.5. A doppler shift, corresponding to \textit{Cluster~A}'s recession velocity \citep[990~km~s$^{-1}$;][]{M15}, is applied to all the spectra. Synthetic magnitudes for the spectra are calculated with the PYSYNPHOT package\footnote{\url{https://pysynphot.readthedocs.io}}.

We try to fit the observed SED of \textit{Cluster~A} with the synthetic ones. In doing this, we adopt a Gaussian prior for the distance modulus, $\mu$, based on the reported distance of 14.7~$\pm$~0.6~Mpc (i.e. $\mu$~=~30.84~$\pm$~0.04). A Gaussian prior is also used for the host galaxy's interstellar reddening based on the value derived by \citetalias{M15} from sodium absorption [$E(B-V)_{\rm host}$~=~0.67~mag], for which we assume a typical uncertainty of 0.05~mag. Flat priors are used for all the other parameters. Figures~\ref{post1.fig} and \ref{post2.fig} display corner plots of the marginalised posterior probability distributions; a best-fitting spectrum is shown in Fig.~\ref{spec.fig}.

The star cluster has a mass of log($M$/$M_\odot$)~=~5.38$^{+0.17}_{-0.12}$ and an age of log($\tau$/yr)~=~7.30$^{+0.07}_{-0.06}$ (i.e. $\tau$~=~20.0$^{+3.5}_{-2.6}$~Myr). We derive a total-to-selective extinction ratio of $R_V$~=~2.6$^{+0.6}_{-0.4}$, which seems to be smaller but still consistent with an average Galactic value of $R_V$~=~3.1. The SED fitting suggests that $f_{\rm rem}$~=~69--95\% of the ionising photons are interacting with the surrounding gas. Thus, the photon leakage effect is non-negligible despite the large uncertainty in $f_{\rm rem}$. The ionisation parameter ($U$) and the hydrogen density ($n_{\rm H}$) cannot be tightly constrained; detailed spectroscopic analysis is required to constrain these two parameters. However, they have very small effects on the determination of the star cluster's age (e.g. Fig.~\ref{post2.fig}).

\citetalias{M15} derived an age of 30--300~Myr by comparing the pre-explosion \textit{BVR} photometry and H$\alpha$ luminosity with BPASS models. Our age estimate has a much smaller uncertainty since we have used more observations that were carried out in recent years. These observations cover a longer wavelength range from the ultraviolet \textit{F275W} band to the near-infrared \textit{F814W} band. Considering the possible contribution from ejecta-CSM interaction, we have very conservatively assume that all post-explosion observations can only provide upper limits for \textit{Cluster~A}'s brightness. Despite this, the upper limits are still very useful in constraining the cluster's SED. For example, the UV upper limits exclude models that are too young and thus too UV-bright, and the \textit{F814W} upper limits rule out old models with strong emission at long wavelengths. Moreover, the post-explosion observations are able to spatially resolve \textit{Cluster~A} and its nearby sources that are confused in the lower-resolution pre-explosion images. This allows us to decontaminate the \textit{WFPC2/F658N} photometry and measure the H$\alpha$ emission of \textit{Cluster~A} more accurately (see Section~\ref{phot.sec} and Appendix~\ref{line.sec}). H$\alpha$ is powered by the ionising photons from massive stars that ionise the surrounding gas. Thus, the bright H$\alpha$ emission also constrains \textit{Cluster~A} to have a young age.

\section{Pre-SN Evolution of the Progenitor System}
\label{progenitor.sec}

\subsection{Summary of Observations}
\label{obs.sec}

\begin{figure*}
\centering
\includegraphics[scale=0.23, angle=0]{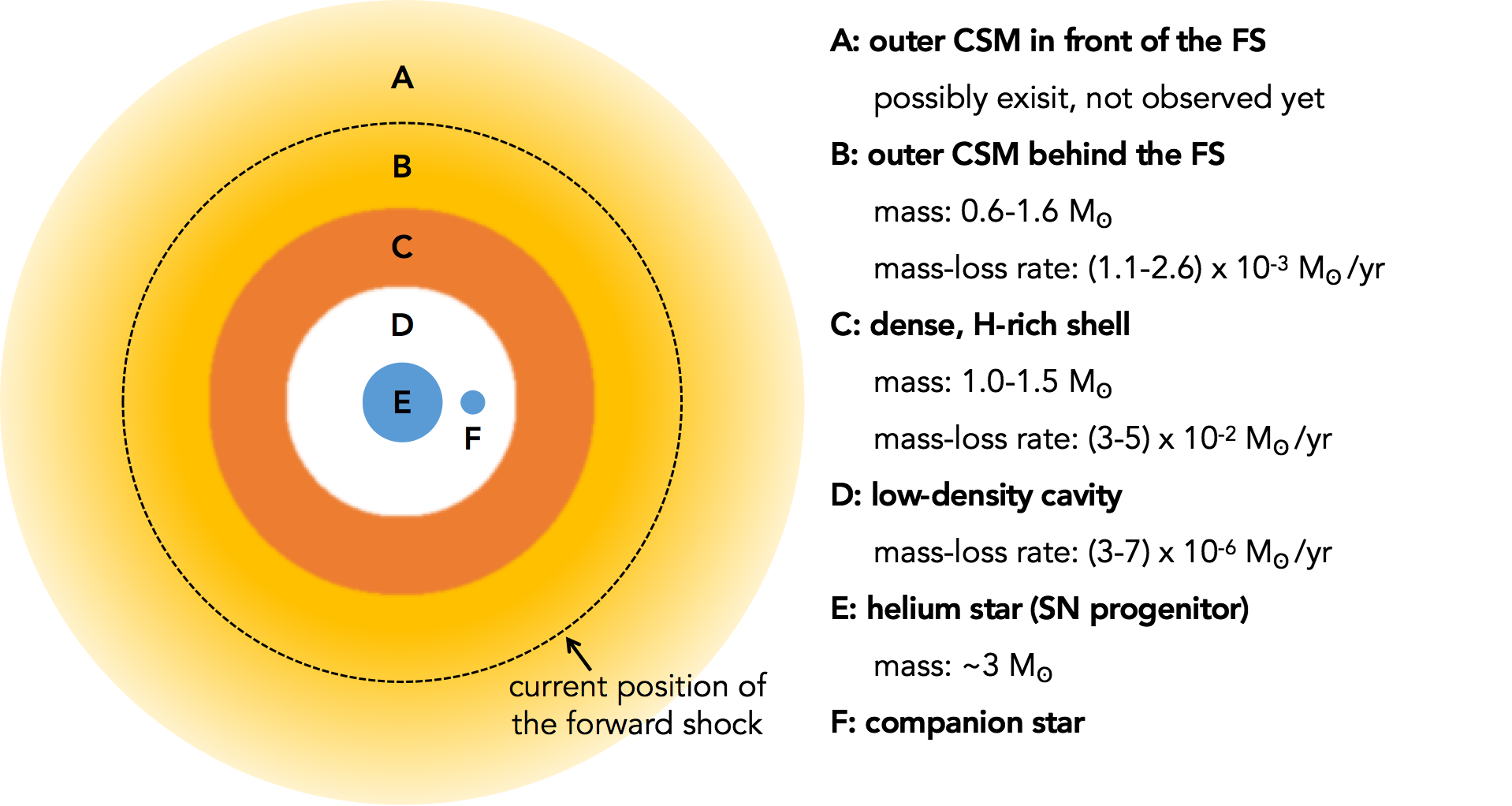}
\caption{A schematic plot of SN~2014C's progenitor and its CSM (not to scale) just before core collapse. The dashed circle shows the forward shock (FS) position at the time of the most recent observations by \citetalias{T19} (1920.8 days post-maximum).}
\label{progenitor.fig}
\end{figure*}

\begin{figure*}
\centering
\includegraphics[scale=0.25, angle=0]{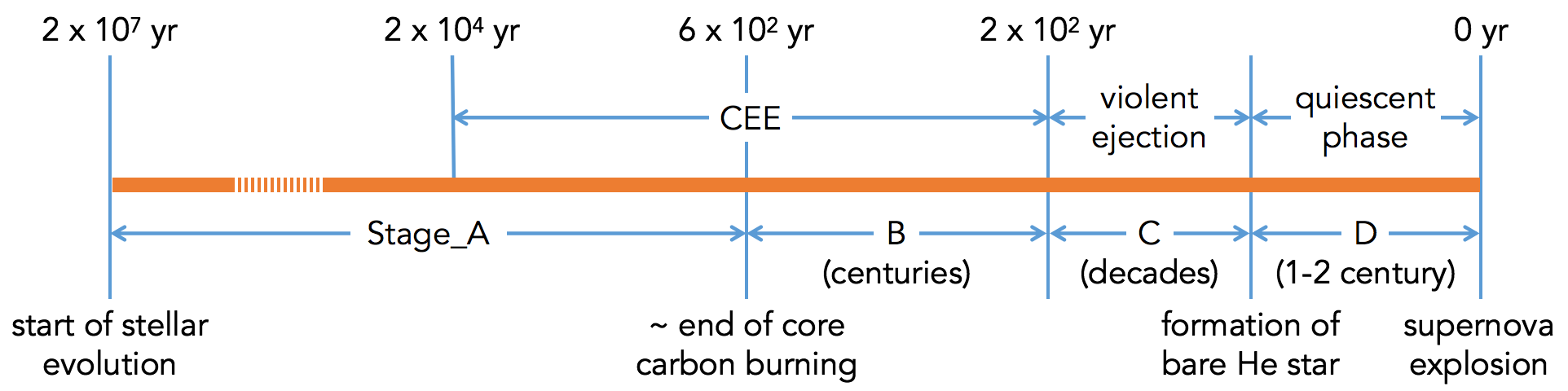}
\caption{Timeline of the evolution of SN~2014C's progenitor, if its early-stage evolution follows that of Model \textit{bpass1} or \textit{bpass2}. At the top of the plot, the timescales correspond to the time before SN explosion.}
\label{time.fig}
\end{figure*}

\begin{figure*}
\centering
\includegraphics[scale=0.25, angle=0]{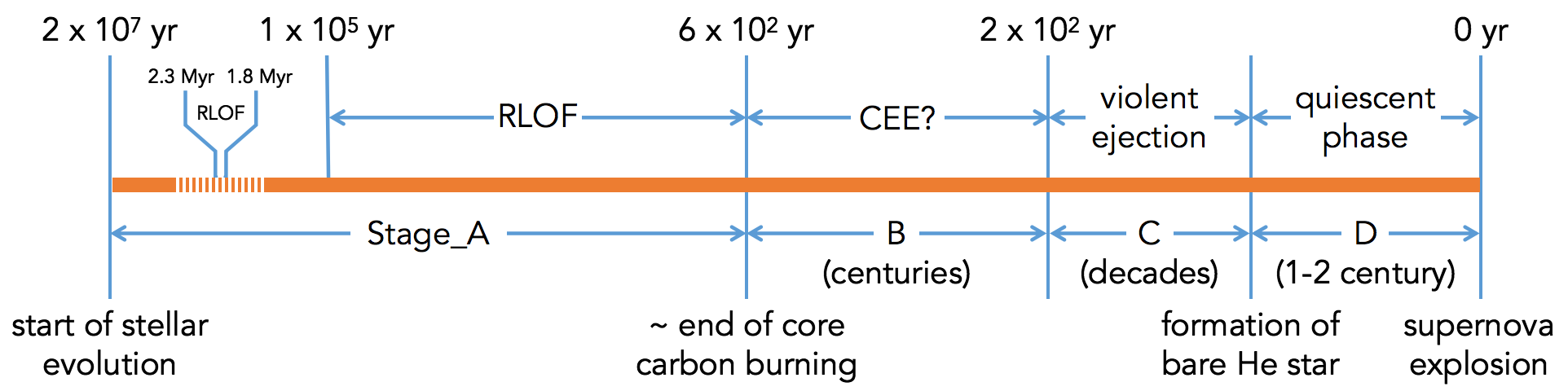}
\caption{Same as Fig.~\ref{time.fig}, but if the early-stage evolution of SN~2014C's progenitor follows that of Model \textit{bpass3}.}
\label{time3.fig}
\end{figure*}

The age of \textit{Cluster~A} corresponds to the lifetime of SN~2014C's progenitor and thus has important implications if the SN progenitor is coeval with its host star cluster. If we assume that the SN progenitor was a single star (or was in a non-interacting binary and ``effectively single"), the lifetime of log($\tau$/yr)~=~7.30$^{+0.07}_{-0.06}$ (i.e. $\tau$~=~20.0$^{+3.5}_{-2.6}$~Myr) corresponds to an initial mass  of $M_{\rm ini}$~=~11.7$^{+1.1}_{-1.0}$~$M_\odot$ (according to the PARSEC v1.2S stellar evolutionary tracks; \citealt{Bressan2012}). However, a single star of this initial mass should still retain a massive hydrogen envelope before it explodes as a hydrogen-rich Type~IIP SN \citep{Smartt2009}. This is inconsistent with SN~2014C; thus, the progenitor star must come from an interacting binary system. In this section we try to explore what kind of binary system can give rise to SN~2014C. To do this, it is important to summarise the observed properties of the progenitor star and the CSM (a schematic plot  is shown in Fig.~\ref{progenitor.fig}):

(1) The progenitor star has a lifetime of log($\tau$/yr)~=~7.30$^{+0.07}_{-0.06}$ (i.e. $\tau$~=~20.0$^{+3.5}_{-2.6}$~Myr; see Section~\ref{property.sec}).

(2) Since SN~2014C exhibited typical Type~Ib spectra at early times, the hydrogen envelope of its progenitor star should be almost completely removed just before core collapse (a small amount of hydrogen may still be left as the near-maximum light spectrum shows an extended, high-velocity H$\alpha$ absorption feature; \citetalias{M15}).

(3) By fitting to its bolometric light curve, \citetalias{M17} find that SN~2014C has a $^{56}$Ni mass of $M_{\rm Ni}$~=~0.15~$\pm$~0.02~$M_\odot$, ejecta mass $M_{\rm ej}$~=~1.7~$\pm$~0.2~$M_\odot$, and explosion kinetic energy $E_{\rm k}$~=~(1.8~$\pm$~0.3)~$\times$~10$^{51}$~erg. The explosion parameters are very typical of Type~Ib/c SNe \citep[see e.g.][]{Lyman2016}. In particular, the low ejecta mass suggests that the helium core of the progenitor star was only $M_{\rm He}$~$\sim$~3~$M_\odot$ if the compact remnant is a neutron star of $\sim$1.4~$M_\odot$ (However, we caution that the derived ejecta mass may vary with the adopted opacity value in the light curve fitting; see, e.g., the discussion in Section~5.7 of \citealt{Maund2018}).

(4) \citetalias{M17} did not detect any X-ray emission during the first $\sim$20~days since the explosion of SN~2014C. The X-ray non-detection suggests that its progenitor star was located in a low-density cavity out to a distance of at least (0.8--2)~$\times$~10$^{16}$~cm. The cavity corresponds to a low mass-loss rate of the progenitor system of $<$~(3--7)~$\times$~10$^{-6}$~$M_\odot$~yr$^{-1}$ for a wind velocity of $v_{\rm w}$~=~1000~km~s$^{-1}$. Thus, the progenitor star did not suffer massive eruptions within 7($v_{\rm w}$/1000~km~s$^{-1}$) years before core collapse. The ejecta-CSM interaction was very weak when the SN ejecta propagated through the low-density cavity; this is why SN~2014C did not exhibit any intermediate-width hydrogen lines at early times \citepalias{M15}.

(5) Outside the low-density cavity, there was a dense, hydrogen-rich CSM shell at a distance of 6~$\times$~10$^{16}$~cm with a thickness of 10$^{16}$~cm, a density of 2~$\times$~10$^6$~cm$^{-3}$, and a mass of 1--1.5~$M_\odot$ \citepalias{M17}. Assuming an ejection velocity of 100~km~s$^{-1}$ (which is the upper velocity limit for the unshocked CSM; the unshocked gas is photoionised by the X-ray photons emitted by the shocked gas and thus can produce narrow emission lines; \citetalias{M15}), the shell was formed $\sim$190~years before core collapse and the mass-loss event lasted for $\sim$30~years. Thus, the shell corresponds to a high mass-loss rate of the progenitor system of (3--5)~$\times$~10$^{-2}$~$M_\odot$~yr$^{-1}$, which is much higher than those of line-driven winds \citep{Smith2014araa}. When the SN ejecta reached this shell at late times, the strong ejecta-CSM interaction produced prominent intermediate-width H$\alpha$ line emission, strong radio and X-ray radiation, and a slower decline rate in the light curve (as kinetic energy was converted into radiative energy; \citetalias{M15}, \citetalias{M17}, \citealt{A17}, \citealt{B18}).

(6) SN~2014C shows on-going ejecta-CSM interaction until 5 years post explosion \citepalias{T19}. Detailed light curve modelling suggests that there is an extended CSM component with a distinct density profile ($\rho$~$\propto$~$r^{-2}$) outside the dense shell. This outer CSM component reaches a distance of 2~$\times$~10$^{17}$~cm and has a mass of 0.6--1.6~$M_\odot$. The inferred mass-loss rate is (1.1--2.6)~$\times$~10$^{-3}$~$M_\odot$~yr$^{-1}$, which is lower than that for the dense shell by an order of magnitude but still much higher than most line-driven winds \citep{Smith2014araa}. Assuming an ejection velocity of 100~km~s$^{-1}$, the mass-loss event that formed this outer CSM component started at least $\sim$630~years before core collapse and lasted until the ejection of the dense shell. The density profile is typical for wind-driven mass loss; however, other types of mass loss can also generate the same density profile as long as the mass-loss rate is quasi-steady \citep[e.g.][]{Deschamps2013, Deschamps2015}. Following Fig.~\ref{progenitor.fig}, we refer to this CSM component as \textit{CSM\_B}.

(7) The total mass of the progenitor star's helium core and the CSM mentioned above is only 4.7--6.2~$M_\odot$, which is smaller than the lower mass limit for a star that can undergo core collapse. This means that the progenitor star must have experienced significant mass loss at even earlier stages, during which it lost a considerable part of its hydrogen envelope. The stripped mass may have been accreted by a binary companion or have formed CSM located at further distances that have not yet been reached by the SN's forward shock (thus, this CSM has not been excited to produce any observable features such as radio/X-ray emission and intermediate-width optical lines until the most recent observations by \citetalias{T19} at 1920.8 days post-maximum). We shall refer to this possible outmost CSM as \textit{CSM\_A}.

In summary, the progenitor star has experienced a complicated evolution with time-varying mass loss. Following the schematic plot of Fig.~\ref{progenitor.fig}, we divide the progenitor star's pre-SN evolution into four stages (designated as \textit{Stage\_A} to \textit{Stage\_D}), during which the different CSM components were formed. Here \textit{Stage\_A} refers to the period from the beginning of stellar evolution until the onset of \textit{Stage\_B}. Since \textit{CSM\_B} were formed during the last few centuries of the progenitor, \textit{Stage\_A} lasted until approximately the end of core carbon burning \citep{Woosley2002}. Next we try to infer the progenitor system's evolution during the four stages. The timelines of two possible scenarios are provided in Figs.~\ref{time.fig} and \ref{time3.fig}.

\subsection{\textit{Stage\_A}}
\label{stageA.sec}

\begin{table*}
\label{bpass.tab}
\center
\caption{BPASS models. Column~1: model ID. Columns~2--4: initial mass of the primary star, initial secondary-to-primary mass ratio, and initial orbital period. Column~5: lifetime of the primary star. Column~6: mass of the residual hydrogen envelope at the end of the simulation. Column~7: compact remnant mass of the primary star if it undergoes a SN explosion of typical explosion energy of 10$^{51}$~erg. Column~8: helium core mass at the end of the simulation. Column~9: helium core mass minus the compact remnant mass, which should be the ejecta mass of SN~2014C (the residual hydrogen envelope was removed after the end of the simulation and just before core collapse).}
\begin{tabular}{ccccccccc}
\hline
\hline
model ID & $M^1_{\rm ini}$/$M_\odot$ & $q_{\rm ini}$ & log($P_{\rm ini}$/days) & log($\tau$/yr) & $M_{\rm env}$/$M_\odot$ & $M_{\rm rem}$/$M_\odot$ & $M_{\rm He}$/$M_\odot$ & ($M_{\rm He}$~$-$~$M_{\rm rem}$)/$M_\odot$ \\
\hline
\textit{bpass1} & 11.0 & 0.2 & 3.0 & 7.35 & 2.02 & 1.46 & 3.26 & 1.80 \\
\textit{bpass2} & 11.0 & 0.3 & 3.0 & 7.35 & 2.83 & 1.45 & 3.26 & 1.81 \\
\textit{bpass3} & 11.0 & 0.9 & 2.8 & 7.35 & 1.83 & 1.45 & 3.15 & 1.70 \\
\hline
\end{tabular}
\end{table*}

\begin{figure*}
\centering
\includegraphics[scale=0.75, angle=0]{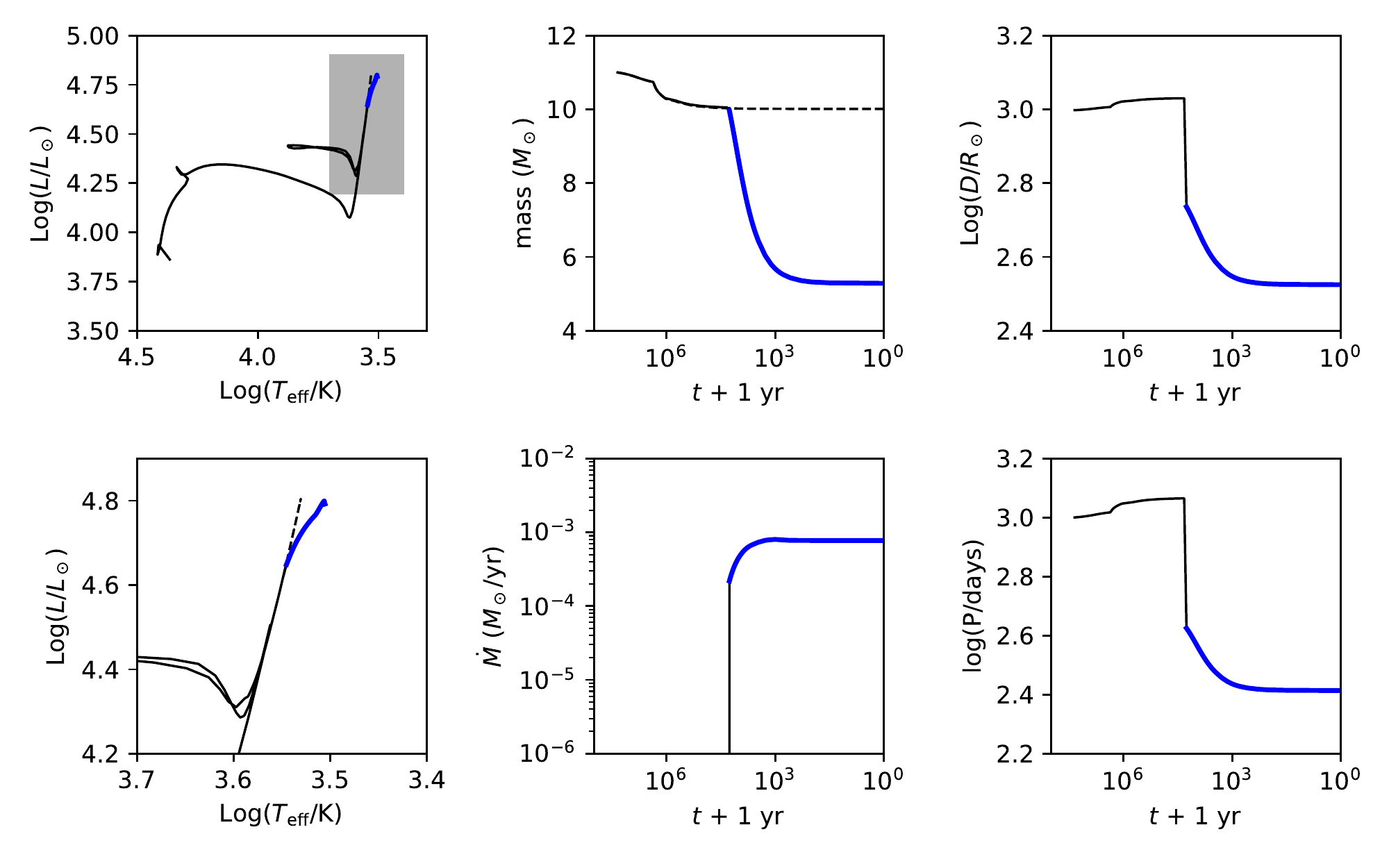}
\caption{Model \textit{bpass1}. The upper-left panel shows the primary star's evolutionary track in the Hertzsprung–Russell diagram, the grey-shaded part of which is zoomed and displayed in the lower-left panel. The middle panels show the evolution of the primary star's mass and mass-loss rate, and the right panels are the evolution of the binary system's orbital separation and period. In the middle and right panels, the horizontal axis corresponds to the time before the end of core carbon burning plus 1 year. The coloured, thick parts of the curves are when the mass transfer is happening between the binary stars. For comparison, the dashed lines correspond to a single star with the same initial mass of the primary star (i.e. 11 $M_\odot$).}
\label{bpass1.fig}
\end{figure*}

\begin{figure*}
\centering
\includegraphics[scale=0.75, angle=0]{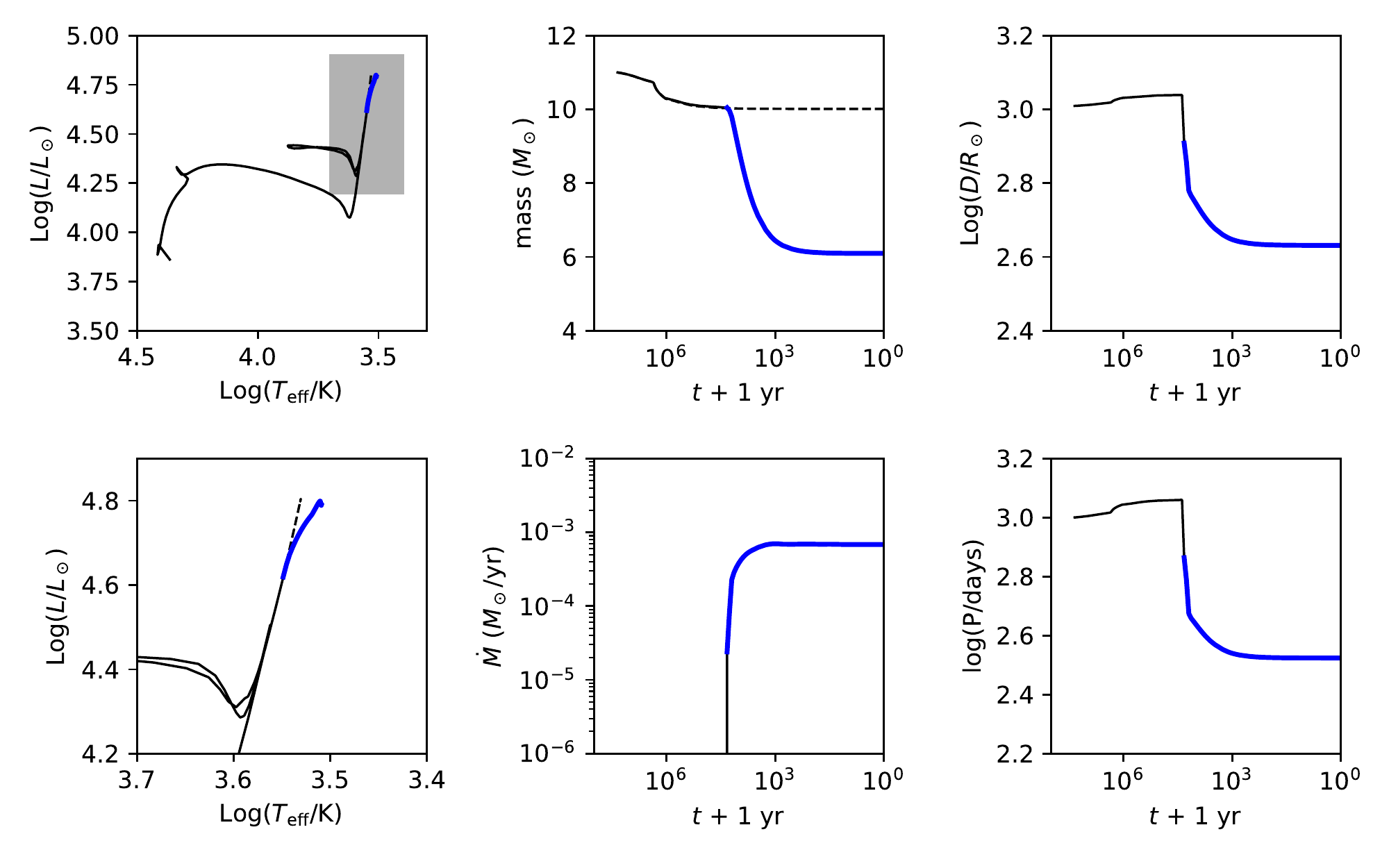}
\caption{Same as Fig.~\ref{bpass1.fig} but for Model \textit{bpass2}.}
\label{bpass2.fig}
\end{figure*}

\begin{figure*}
\centering
\includegraphics[scale=0.75, angle=0]{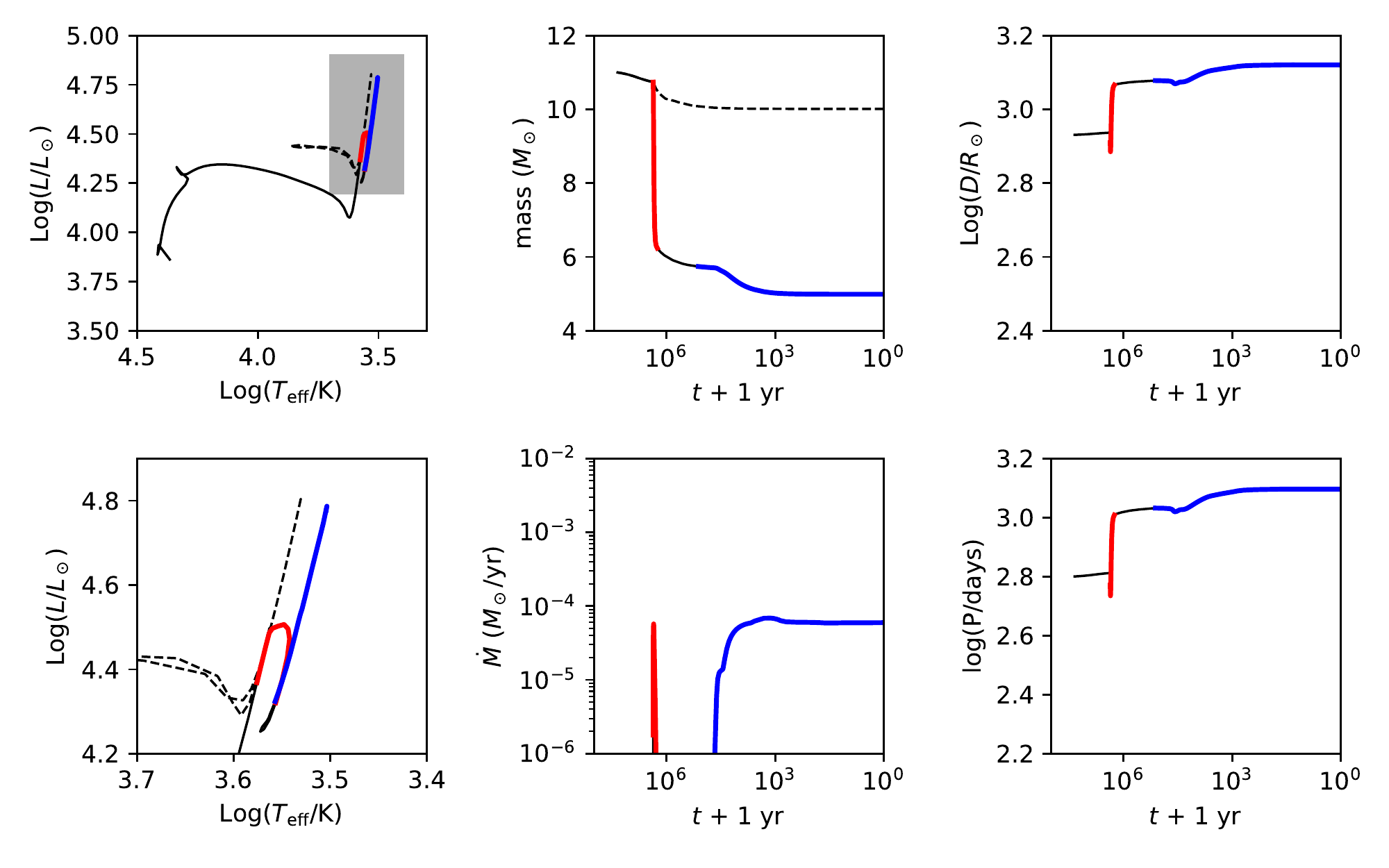}
\caption{Same as Fig.~\ref{bpass1.fig} but for Model \textit{bpass3}.}
\label{bpass3.fig}
\end{figure*}

BPASS \citep{bpass.ref} provides a grid of binary stellar evolution models spanning a wide range of parameters. The stellar evolution modesl are calculated until the end of core carbon burning, or neon ignition for the most massive stars. Thus, they are useful to investigate the \textit{Stage\_A} evolution of SN~2014C's progenitor system and we try to search for models with an end point meeting the following criteria:

(1) lifetime of the primary star is consistent with that of SN~2014C's host star cluster, i.e. log($\tau$/yr)~=~7.30$^{+0.07}_{-0.06}$ (i.e. $\tau$~=~20.0$^{+3.5}_{-2.6}$~Myr);

(2) the He core mass ($M_{\rm He}$) is equal to that of the compact remnant (derived by BPASS for a typical SN explosion energy of 10$^{51}$~erg) plus SN~2014C's ejecta mass of 1.7~$\pm$~0.2~$M_\odot$ \citepalias{M17};

(3) the primary star still has a residual hydrogen envelope with a mass of $M_{\rm env}$~=~1.6--3.1~$M_\odot$, which will be lost to form \textit{CSM\_B} \citepalias{T19} and the dense shell at later stages (\citetalias{M15}; \citealt{A17}; \citetalias{M17}).

Three models from BPASS can satisfy the above criteria, the parameters of which are listed in Table~3. All three models have a primary star with an initial mass of 11~$M_\odot$. Models \textit{bpass1} and \textit{bpass2} have the same initial orbital period of log($P_{\rm ini}$/days)~=~3.0 and differ only in the initial secondary-to-primary mass ratio ($q_{\rm ini}$~=~0.2 for \textit{bpass1} and $q_{\rm ini}$~=~0.3 for \textit{bpass2}). Their evolution is also very similar. Figures~\ref{bpass1.fig} and \ref{bpass2.fig} show the primary star's evolutionary track in the Hertzsprung–Russell diagram, the change in the primary star's mass and mass-loss rate, and the evolution of the orbital separation and period of the binary system. Models \textit{bpass1} and \textit{bpass2} both undergo \textit{Case-C} mass transfer (i.e. the mass-donor star has already completed core helium burning when the mass transfer occurs) at very late times. The mass transfer is via CE ejection, starting from 2~$\times$~10$^4$ years before the end of core carbon burning. The mass-loss rate reaches at most close to $\sim$10$^{-3}$~$M_\odot$~yr$^{-1}$ and several solar masses of the hydrogen envelope is removed until the end of the calculation.

Model~\textit{bpass3} has a larger initial secondary-to-primary mass ratio of $q_{\rm ini}$~=~0.9 and a shorter initial orbital period of log($P_{\rm ini}$/days)~=~2.8. Unlike Models \textit{bpass1} and \textit{bpass2}, Model \textit{bpass3} undergoes \textit{Case~BC} mass transfer: the first episode of mass transfer occurs when the primary star is undergoing hydrogen shell burning while the second episode starts when the star has finished core helium burning. Both mass-transfer episodes are via RLOF and are nearly conservative, i.e. all mass lost from the primary star is accreted by the secondary. However, the mass-loss rate is $<$~10$^{-4}$~$M_\odot$~yr$^{-1}$ and much lower than those of models \textit{bpass1} and \textit{bpass2}.

\subsection{\textit{Stage\_B}}
\label{stageB.sec}

\subsubsection{Models \textit{bpass1} and \textit{bpass2}}

It would be interesting to explore the further evolution of the progenitor system at later stages. For Models \textit{bpass1} and \textit{bpass2}, the mass-loss rate of the CE ejection at the end of core carbon burning [(0.7--0.8)~$\times$~10$^{-3}$~$M_\odot$~yr$^{-1}$] is very close to that required for SN~2014C at \textit{Stage\_B} [(1.1--2.6)~$\times$~10$^{-3}$~$M_\odot$~yr$^{-1}$; \citetalias{T19}]. Thus, it is very likely that the CE evolution continues to \textit{Stage\_B} after the end of core carbon burning. In other words, \textit{CSM\_A} and \textit{CSM\_B} in Fig.~\ref{progenitor.fig} are the CE that was ejected by the binary interaction. In the future, we may be able to see ongoing ejecta-CSM interaction as the forward shock propagates to further distances into the ejected material.

\subsubsection{Model \textit{bpass3}}

For Model \textit{bpass3}, the mass-loss rate for its second episode of mass transfer is only 6~$\times$~10$^{-5}$~$M_\odot$~yr$^{-1}$ at the end of core carbon burning. Moreover, the secondary star accretes almost all the material lost by the primary, leaving little to form a CSM (in this case, \textit{CSM\_A} in Fig.~\ref{progenitor.fig} may not exist at all). Thus, if SN~2014C's progenitor has followed the evolution of Model \textit{bpass3}, we need some mechanism to increase the mass-loss rate by a factor of 20--40 and to form a dense CSM at \textit{Stage\_B}. For example, the envelope of RSGs may become dynamically unstable and develop large-amplitude pulsations \citep{Yoon2010}; the fast envelope expansion may lead to a transition in the form of mass transfer from RLOF to CE evolution. However, more detailed modelling is needed to precisely understand the progenitor's late-time evolution.

\subsection{\textit{Stage\_C}}
\label{stageC.sec}

\textit{Stage\_B} was followed by a violent mass ejection (i.e. \textit{Stage\_C}) in which 1--1.5~$M_\odot$ of the last hydrogen layer (or 24--32\% of the star's current mass) was lost within several decades to form the dense shell around SN~2014C \citepalias{M15, M17}. The mass-loss rate abruptly increased by an order of magnitude to (3--5)~$\times$~10$^{-2}$~$M_\odot$~yr$^{-1}$ and the ejected dense shell has a constant density profile, which is different from that of the outer CSM component (\citetalias{T19}; see also the discussion in Section~\ref{obs.sec}). This suggests that the mass loss at \textit{Stage\_C} may arise from a different mechanism from \textit{Stage\_B}. The mass-loss rate at \textit{Stage\_C} is comparable to those of LBV giant eruptions and the pre-SN eruptions of Type~IIn/Ibn SN progenitors \citep{Smith2017book}. Thus, \textit{CSM\_C} is likely to arise from a violent eruption from the progenitor star that occurred very close in time to core collapse.

\subsection{\textit{Stage\_D}}
\label{stageD.sec}

After the violent ejection at \textit{Stage\_C}, the progenitor star was almost a bare helium core of $\sim$3~$M_\odot$ with very little hydrogen envelope. The low-density cavity around SN~2014C corresponds to a very low mass-loss rate [$<$~(3--7)~$\times$~10$^{-6}$~$M_\odot$~yr$^{-1}$; see the discussion of Section~\ref{obs.sec}], suggesting that the progenitor remained quiescent in the final stage before its final explosion.

However, we cannot exclude the possibility that the progenitor was still subject to wind mass loss at this stage. For low-mass helium stars, the wind mass-loss prescription is still very uncertain, since very few such object have been identified from observations. By extrapolating the empirical prescription for WR stars \citep[e.g.][]{Nugis2000}, \citet{Gotberg2018} find a low mass-loss rate of log($\dot{M}$/$M_\odot$~yr$^{-1}$)~=~$-$6.8 for a helium star of 3.32~$M_\odot$ at solar metallicity. The theoretical prescription of \citet{Vink2017}, however, predicts much weaker winds and that a 3-$M_\odot$ helium star has a mass-loss rate of only log($\dot{M}$/$M_\odot$~yr$^{-1}$)~=~$-$7.82. Despite the large uncertainty, both predictions do not exceed the upper limit for SN~2014C's mass-loss rate inferred from its low-density cavity.

\ \\
\ \\

\begin{figure*}
\centering
\includegraphics[width=0.9\linewidth, angle=0]{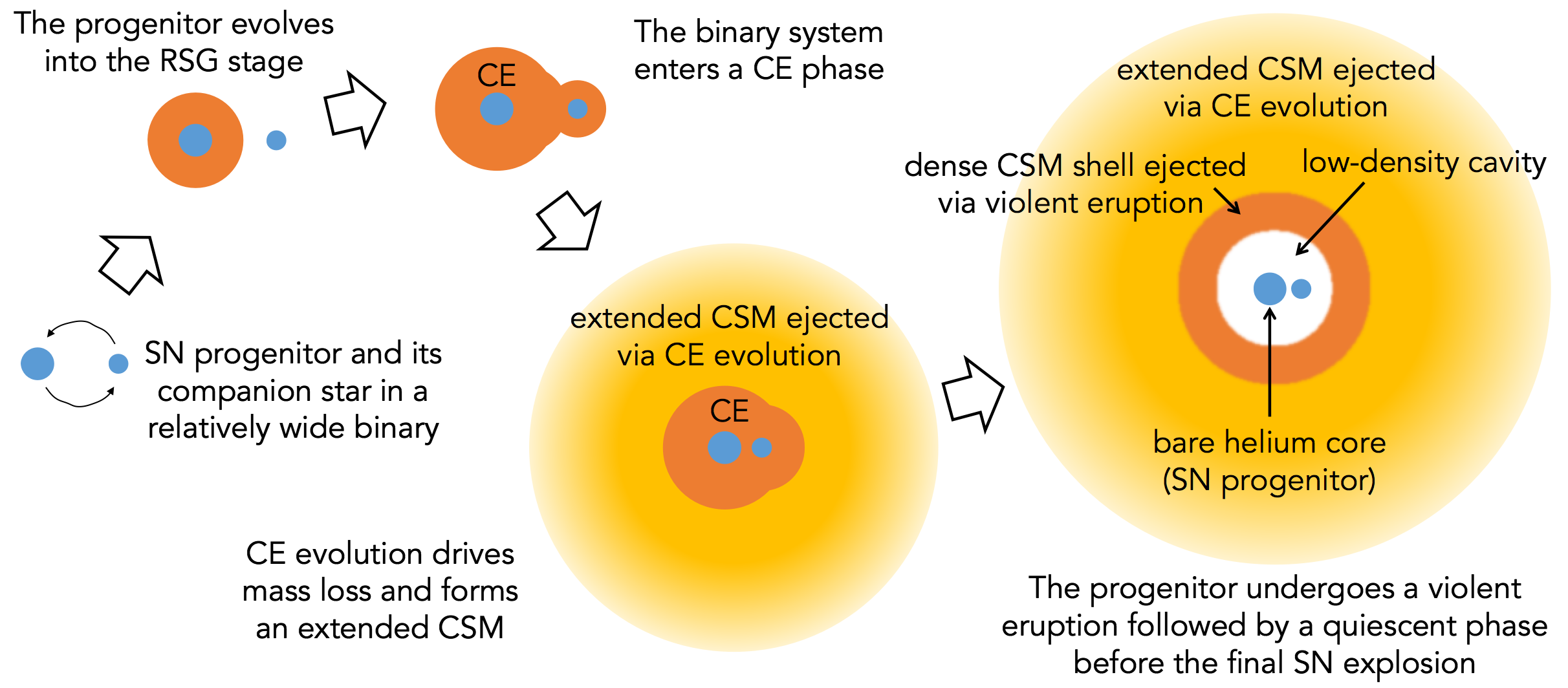}
\caption{A schematic plot of SN~2014C's progenitor system and its pre-SN evolution (based on Models \textit{bpass1} and \textit{bpass2}; for Model \textit{bpass3}, the evolution is only slightly different; see text for details).}
\label{channel.fig}
\end{figure*}

In summary, BPASS models suggest the progenitor of SN~2014C was an 11-$M_\odot$ star in a relatively wide binary system that has experienced a complex mass-loss history (a schematic plot is shown in Fig.~\ref{channel.fig}). Due to the relatively large initial orbital separation ($\sim$10$^{3}$~$R_\odot$), the progenitor underwent a late episode of binary interaction very close to the end of its life. As a result, the ejected material formed an extended CSM component, which had not fully dissipated into the interstellar space when it was caught up by SN~2014C's ejecta. The binary interaction was only able to eject part of the hydrogen envelope before a violent eruption from the progenitor star ejected the remaining part. During the eruption, 1--1.5~$M_\odot$ of the envelope was ejected within just several decades, which formed a dense shell of CSM around the progenitor system. After that, the progenitor star was almost a bare helium core and its mass loss was very weak, if any. This quiescent phase lasted for only 1--2 centuries before the SN explosion and led to the formation of a low-density cavity between the progenitor and the dense CSM shell.

\section{Summary and Discussion}
\label{summary.sec}

In this paper, we report a detailed analysis of the host star cluster of the transforming Type~IIn/Ibn SN~2014C. The analysis is based on \textit{HST} observations, which cover a long wavelength range from the UV to the \textit{F814W} band. The dataset also includes two narrow-band filters centred on the H$\alpha$ and [N~{\small II}] lines. We perform accurate photometry on the host star cluster and, when necessary, removed the contamination from nearby sources. We have reached the following conclusions:

(1) We find that the radiation from SN~2014C was still significant until at least 4 years after its explosion. This is consistent with the recent discovery that SN~2014C has an on-going interaction between the ejecta and extended CSM at far distances \citepalias{T19}. Still, photometry at post-explosion images provides important upper limits for the host star cluster's brightness.

(2) We model the host star cluster's SED with BPASS for the stellar population and CLOUDY for the gaseous nebula. The result suggests that it has an age of log($\tau$/yr)~=~7.30$^{+0.07}_{-0.06}$ (i.e. $\tau$~=~20.0$^{+3.5}_{-2.6}$~Myr) and a mass of log($M$/$M_\odot$)~=~5.38$^{+0.17}_{-0.12}$. The age estimate rules out a single star progenitor for SN~2014C, assuming the progenitor to be coeval with the host star cluster. Thus, SN~2014C is most likely to arise from an interacting binary system.

(3) We further try to infer the initial configuration and the pre-SN evolution of the binary progenitor system. We find that three models from BPASS may be able to account for the early-stage evolution of the progenitor system. All are relatively wide binaries with an initial orbital period of log($P_{\rm ini}$/days)~=~2.8--3.0 and an initial mass of 11~$M_\odot$ for the primary star. They underwent a \textit{Case~C} or \textit{Case~BC} mass transfer. In the former case, the mass transfer was via CE evolution and the ejected CE formed the extended CSM outside the dense shell around SN~2014C.

(4) The progenitor star then experienced a violent eruption, in which 1.0--1.5~$M_\odot$ of its last hydrogen layer was lost within decades to form a dense CSM shell. After the violent eruption, the progenitor star was almost a bare helium star, which stayed quite quiescent (although possibly still had a weak stellar wind) and ready for its final explosion.

Violent eruptions seem to be quite common for a significant number of (but not all) massive stars. The eruptions may be responsible for the dense CSM around Type~IIn/Ibn SNe and are sometimes directly observed as optical outbursts before the terminal SN explosion (e.g. SN~2006jc, \citealt{Pastorello2007}; SN~2009ip, \citealt{Mauerhan2013}; see also \citealt{Ofek2014}). These phenomena suggest that the massive stars may become wildly unstable in the last centuries, decades, or even years before the end of their lives - something that has not been included in standard stellar evolutionary models.

Significant progress has been achieved in the past decade to understand the mechanism for the pre-SN eruptions of massive stars. For example, hydrodynamic waves may be excited by vigorous core convection at late nuclear burning stages; the waves, which propagate outward with a super-Eddington energy flux, can trigger an outburst of mass loss if they are able to reach close enough to the stellar surface \citep{Quataert2012, Shiode2014, Fuller2017, Fuller2018}. Moreover, the vigorous convection may cause finite-amplitude fluctuations in temperature, density, etc., and will be coupled non-linearly with nuclear burning. This may result in unsteady or even explosive nuclear burning that contribute to the mass ejection at late stages \citep{Smith2014convection}. Yet, more studies are still needed to fully understand the pre-SN eruptions.

It is also worth noting that SN~2014C's dense shell corresponds to an extremely high mass-loss rate [(3--5)~$\times$~10$^{-2}$~$M_\odot$~yr$^{-1}$], which is similar to those of LBV giant eruptions. It was believed that such violent eruptions can only occur in stars which are initially more massive than 25~$M_\odot$. Recent observations, however, seem to suggest that eruptions with comparable intensities can also occur in lower-mass massive stars (a thorough discussion can be found in Section~6.3.2 of \citetalias{S20}). In particular, \citet{Shivvers2017} and \citetalias{S20} showed that the Type~Ibn SNe 2006jc and 2015G may arise from lower-mass, interacting binary systems; yet, their progenitors underwent violent mass loss just before core collapse with similar characteristics of LBV giant eruptions (for SN~2006jc, the eruption was also observed as an optical outburst in 2004, two years before its final explosion; \citealt{Pastorello2007}). This led \citetalias{S20} to conclude that lower-mass massive stars (8~$<$~$M_{\rm ini}$~$<$~25~$M_\odot$) can also experience violent pre-SN eruptions that resemble LBV giant eruptions, if their envelopes are stripped in interacting binaries. The theoretical work of \citet{Fuller2017} and \citet{Fuller2018} also suggests that the removal of the envelope may aid the occurrence of pre-SN eruptions as in this case the convection-excited waves are easier to reach and trigger an outburst at the stellar surface.

SN~2014C may be the third example to support such a conclusion since its progenitor had an initial mass of only $\sim$11~$M_\odot$ and was stripped by its companion. One difference from SNe~2006jc and 2015G is that, at the time of the eruption, SNe~2006jc's and 2015G's progenitors were almost bare helium stars while SN~2014C's progenitor was only partially stripped and still had a residual hydrogen envelope. This residual hydrogen envelope accounted for 24--32\% of its total mass and was almost completely ejected within a few decades. Thus, the eruption was surprisingly violent for its mass and had a profound effect on the end product.

Another difference is also considerable between SN~2014C and other ``normal" Type~IIn/Ibn SNe. For SN~2014C, the dense shell is located at a remote distance from the progenitor and its interaction with the SN ejecta started at very late times \citep[$\sim$200~days after explosion;][]{A17}. In contrast, many interacting SNe show strong ejecta-CSM interaction almost immediately after the SN explosion. This suggests that SN~2014C's dense shell was ejected a much longer time before the progenitor's core collapse than those of the ``normal" Type~IIn/Ibn SNe. Such a difference may be related to different nuclear burning stages, stellar structures and/or energy sources for their mass loss. \citetalias{M17} studied a sample of Type~Ib/c SNe and found that $\sim$10\% of them show a radio re-brightening at late times. Thus, it might be more common than previously thought for otherwise ``normal" SNe to have CSM at remote distances just like SN~2014C. Many such SNe may have eluded our attention since they are not monitored continuously out to years or decades.

\citetalias{M17} also discussed the formation of SN~2014C's dense shell through a \textit{Case~C} binary interaction \textit{or} through an eruption of its progenitor. In this paper, we find that both processes may have taken place, but they occurred at different evolutionary stages and were responsible for the formation of different CSM components. The ejected CE may be much more extended than the current position of the SN's forward shock. Thus, we predict that SN~2014C may remain bright in the next few years with enduring ejecta-CSM interaction.

In this paper we have assumed that the SN progenitor was the primary star in the system. Yet, we cannot exclude the possibility that the progenitor may be the initially secondary star with a compact companion (which was left after the primary star underwent core collapse). In this case, a merger between the two stars may also be able to eject the CE, trigger the violent mass ejection \citep{Podsiadlowski2006} and induce the subsequent SN explosion \citep{Chevalier2012}. More detailed modelling is needed to fully and accurately understand the pre-SN evolution of SN~2014C's progenitor system.

%%%%%%%%%%%%%%%

\section*{Acknowledgements}

We are grateful to the anonymous referee for his/her comments in improving this paper. N-CS acknowledges the funding support from the Science and Technology Facilities Council. This work is based on observations conducted with the Hubble Space Telescope and the data were retrieved from the data archive at the Space Telescope Science Institute. This work has also made use of the BPASS stellar population models.

\section*{Data Availability}

The data used in this work are all publicly available from the Mikulski Archive for Space Telescope (\url{https://archive.stsci.edu}) and the Hubble Legacy Archive (\url{https://hla.stsci.edu}).

\appendix

\section{Photometry with GALFIT}
\label{galfit.sec}

\begin{figure*}
\centering
\includegraphics[scale=0.55, angle=0]{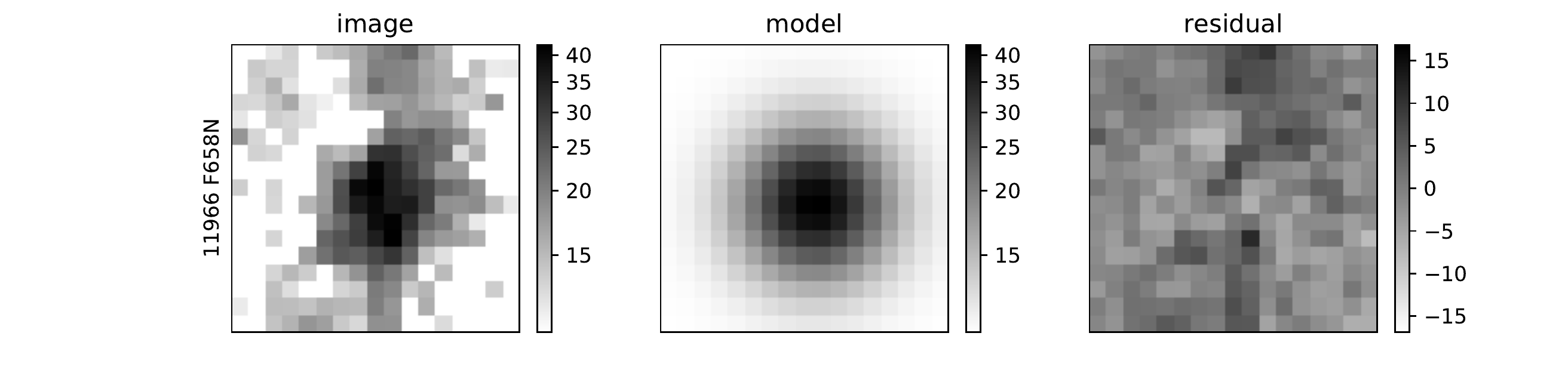}
\includegraphics[scale=0.55, angle=0]{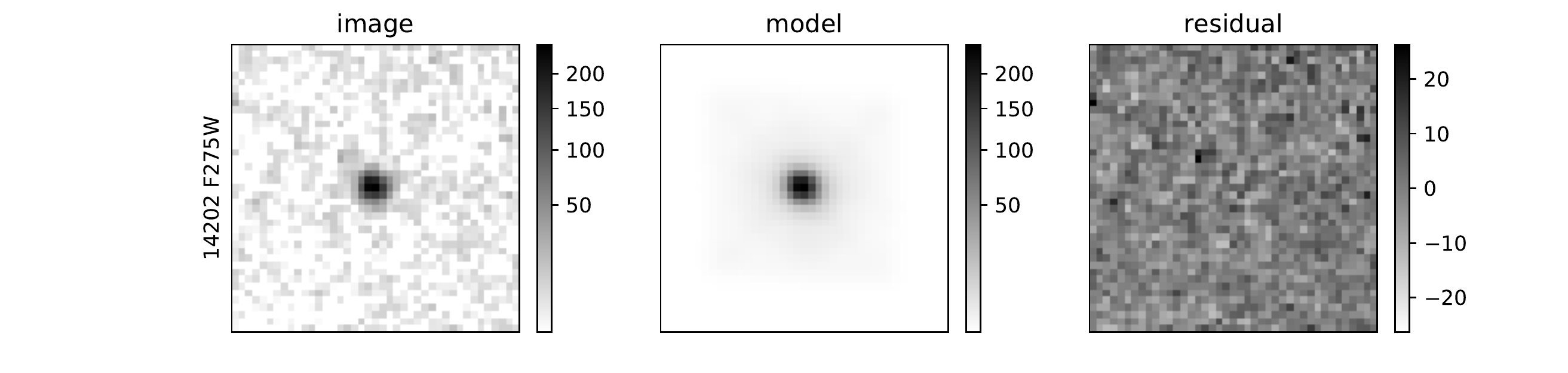}
\includegraphics[scale=0.55, angle=0]{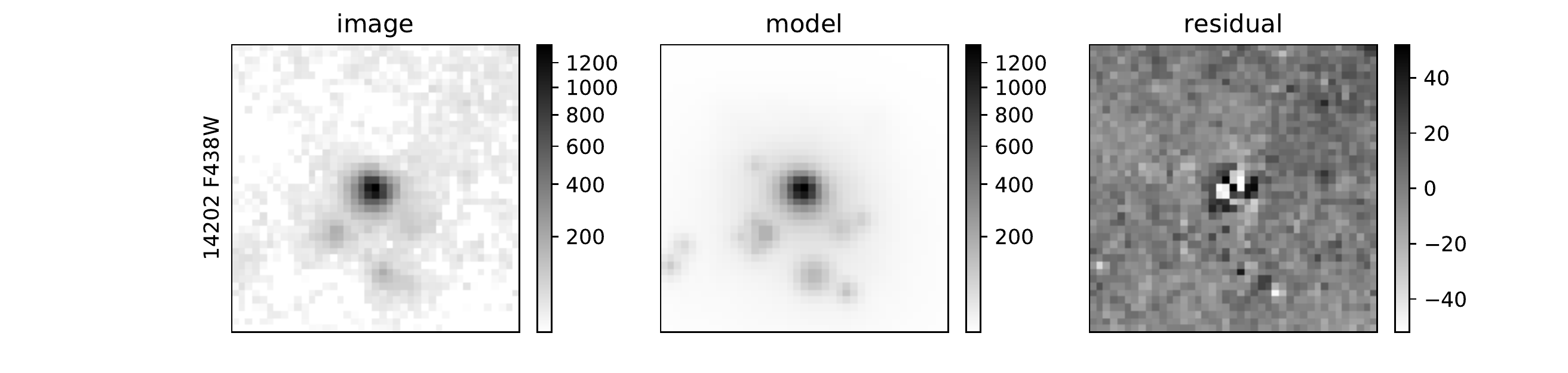}
\includegraphics[scale=0.55, angle=0]{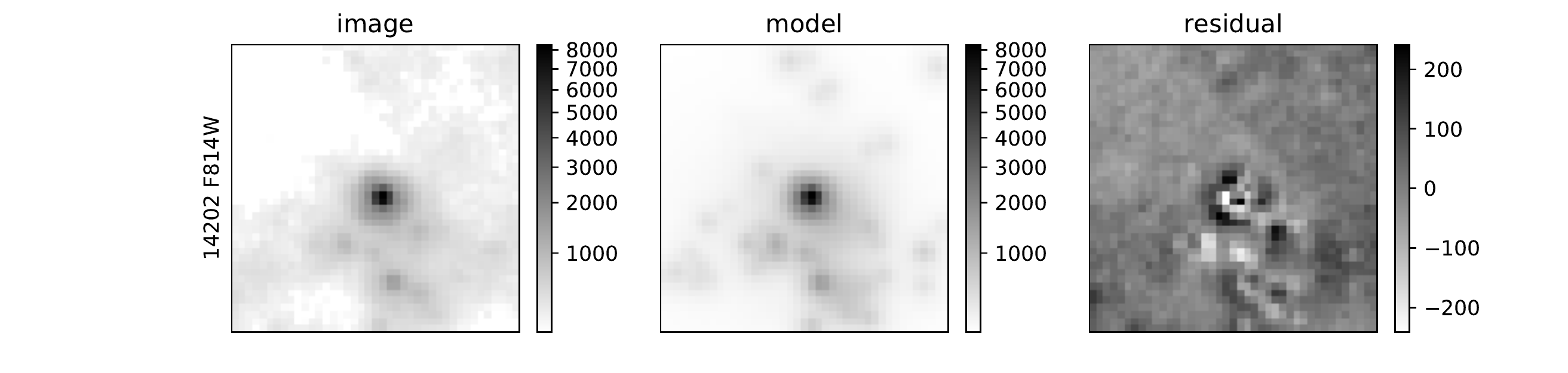}
\includegraphics[scale=0.55, angle=0]{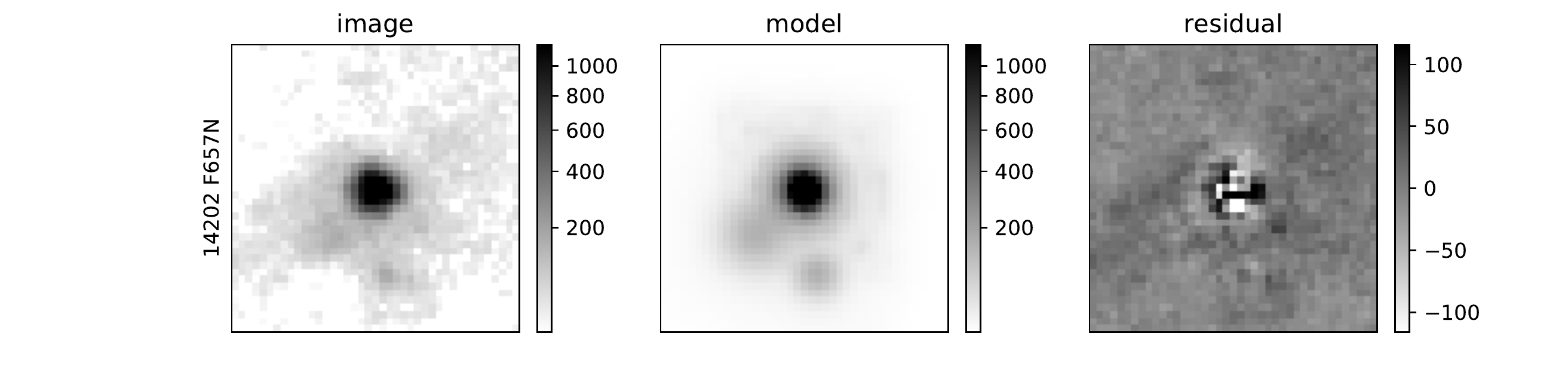}
\includegraphics[scale=0.55, angle=0]{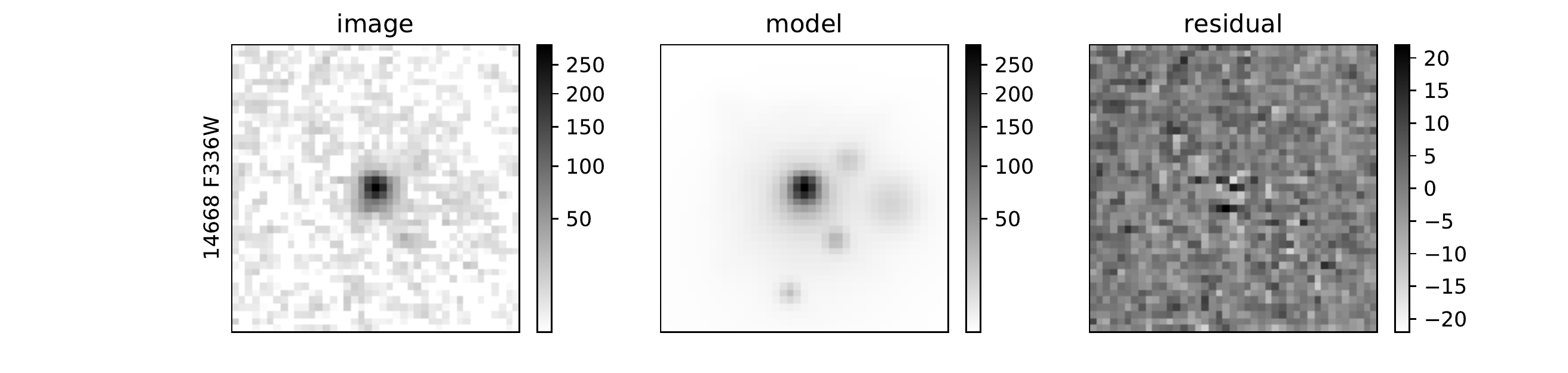}
\caption{Images, models, and the residuals of the star cluster region. The colour bars are in units of analog-to-digital units. The images are centred on \textit{Cluster~A} and aligned with the image axes. Each set of image is labeled with its \textit{HST} program ID and its filter (see Table~1).}
\label{galfit1.fig}
\end{figure*}

\begin{figure*}
\centering
\includegraphics[scale=0.55, angle=0]{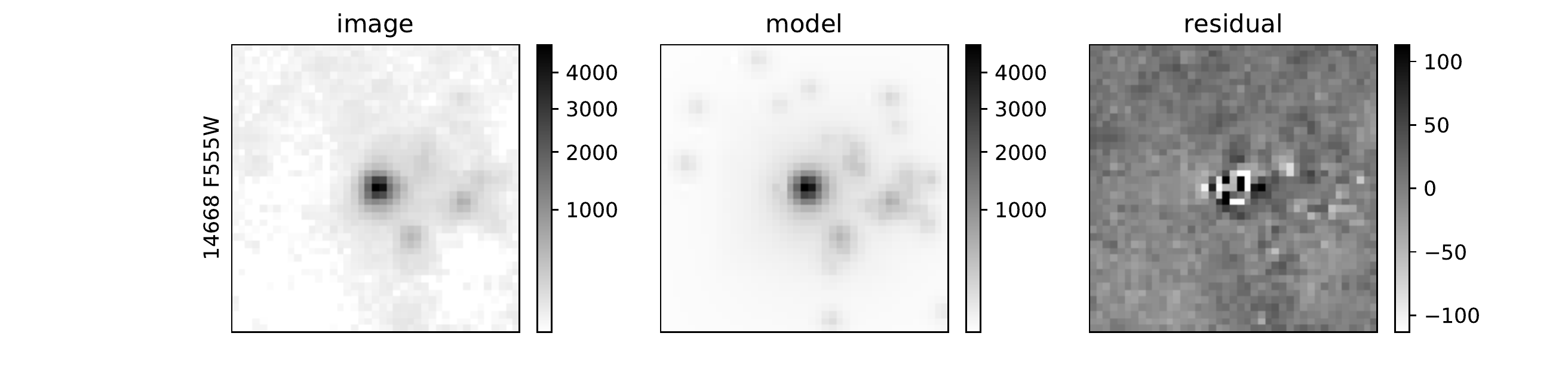}
\includegraphics[scale=0.55, angle=0]{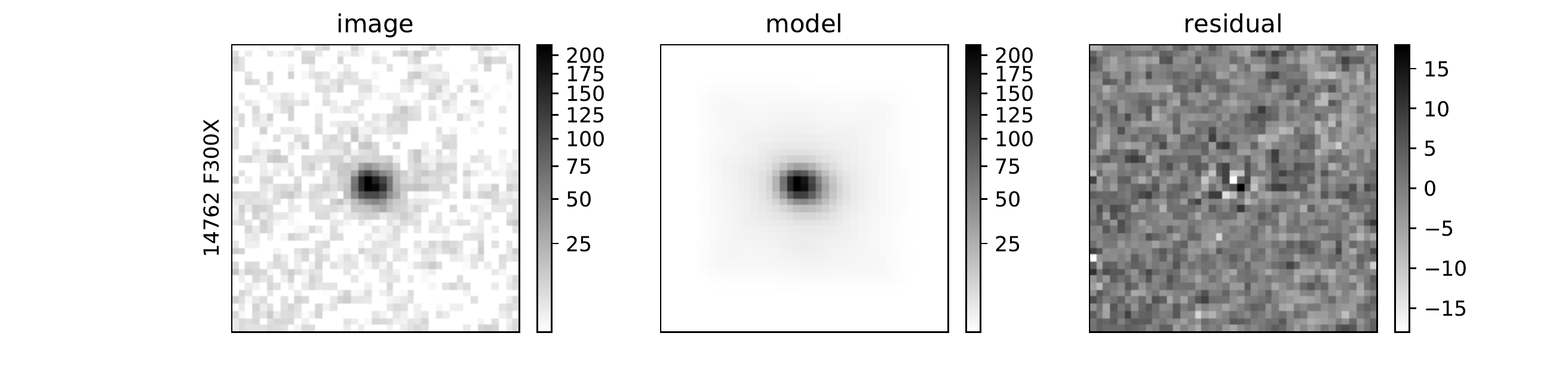}
\includegraphics[scale=0.55, angle=0]{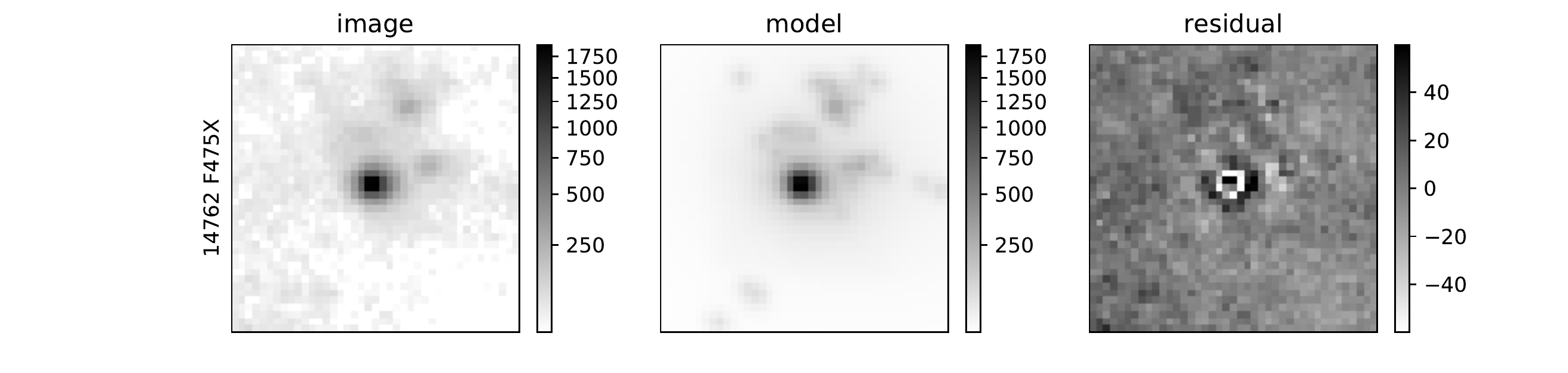}
\includegraphics[scale=0.55, angle=0]{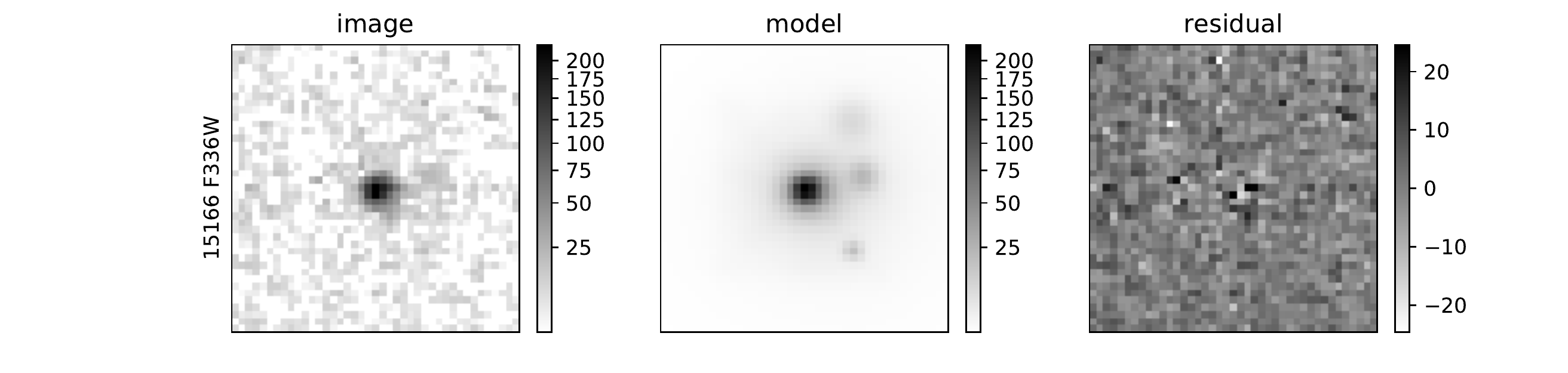}
\includegraphics[scale=0.55, angle=0]{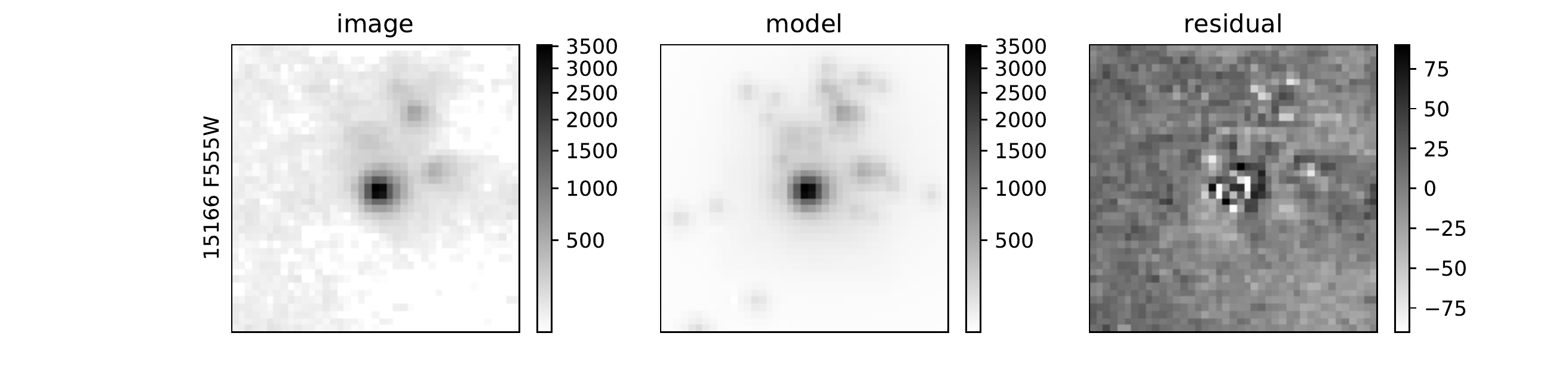}
\includegraphics[scale=0.55, angle=0]{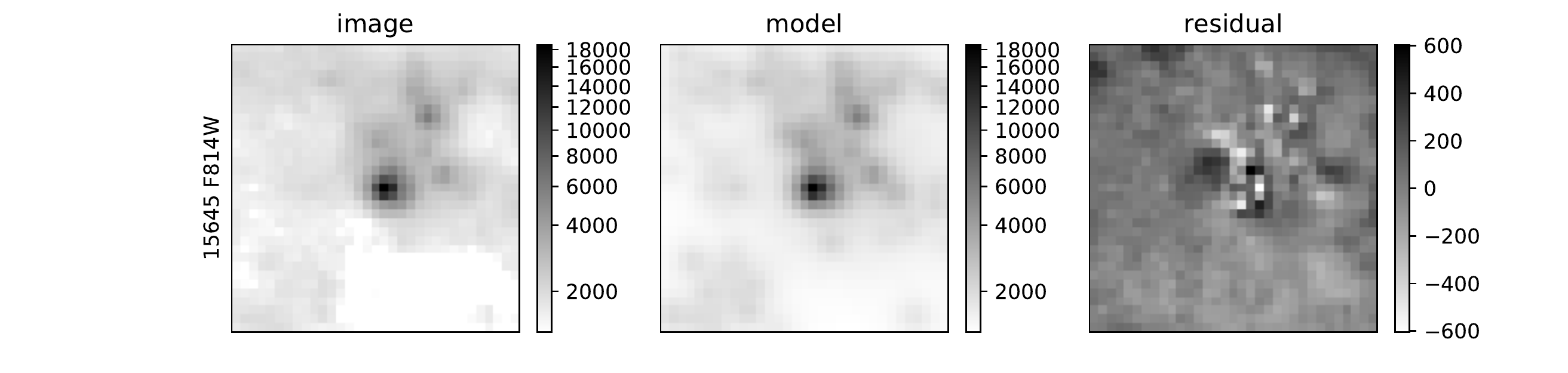}
\caption{Continued}
\label{galfit2.fig}
\end{figure*}

It is not trivial to perform photometry on unresolved star clusters that are more extended than point sources, especially when they are in crowded regions. Aperture photometry with a large aperture radius will unavoidably include the contamination from nearby sources. Alternatively, if one uses a small aperture radius, it is necessary to perform aperture correction to compensate for the light outside the aperture radius. However, the aperture correction for an extended source may be different from that derived from stars. To overcome these difficulties, we use the GALFIT package \citep{galfit1.ref, galfit2.ref} to model the \textit{HST} images of \textit{Cluster~A} and its nearby sources (within 1.6$\arcsec$~$\times$~1.6$\arcsec$ box regions centred on \textit{Cluster~A} and aligned with the image axes; Figs.~\ref{galfit1.fig}, \ref{galfit2.fig}). In this way, the star cluster and its nearby sources can be isolated and their magnitudes can be derived accurately from model fitting. Model PSFs, generated with the TinyTim package, are used in GALFIT.

For the \textit{F658N} band, \textit{Cluster~A} falls on the \textit{WF3} chip of \textit{WFPC2}, which has a poorer spatial resolution compared with the other images taken by \textit{WFC3/UVIS} and \textit{ACS/WFC}. As a result, the emission from \textit{Cluster~A}, \textit{S1} and \textit{S2} cannot be spatially resolved and appears to be a single, extended source. Thus, we model the image with a single Gaussian profile plus a flat sky background.

The \textit{F657N}-band image has a higher spatial resolution and reveals \textit{S1} and \textit{S2} in the vicinity of \textit{Cluster~A}; we use a Gaussian profile to model each of them. For \textit{Cluster~A}, however, we find that a single Gaussian profile does not reproduce its light distribution; instead, it is better described by the sum of two Gaussian profiles with different FWHMs. A flat sky background is also included.

In the broad-band images, the light of \textit{Cluster~A} comes mainly from its stellar component. We model the star cluster with a Moffat profile
\begin{equation}
\Sigma (r) \propto \dfrac{1}{\left[1 + (r/r_c)^2 \right]^{\gamma/2}}
\end{equation}
where $\Sigma$ is the surface brightness, \textit{R} the projected distance to the cluster centre, and $r_c$ the core radius. This profile has a flat core near the centre with a power-law decline at larger radius and can well describe the light profile of young star clusters \citep{Elson1987}. For the nearby stars in the field, we first run the DOLPHOT package \citep{dolphot.ref} to get a stellar catalogue with positions and magnitudes; each star is then modelled with a narrow Gaussian function with a position and magnitude as derived by DOLPHOT and a FWHM of 0.5~pixel (which is equivalently a delta function). For the brightest three stars in each image, we allow their positions, magnitudes, and FWHMs to be adjusted during the fitting. This is because the stars may not be perfect PSFs as they may be confused by other nearby stars or may even be unresolved star clusters. For all the other stars their model parameters are fixed to save computing power. A flat sky background is included.

The broad-band images can be fitted well with the above models, except for the \textit{F814W} band, where large residuals exist at the positions of field stars. The number of field stars are the most numerous in the \textit{F814W}-band images and, as a result, source crowding and confusion is most severe in these images. With a few tests, we find that they can be modelled best if we use larger FWHMs in the Gaussian profiles for the field stars (1.0~pixel for the \textit{F814W}-band image of Program 14202 and 1.5 pixel for that of Program 15645). This effectively uses a wider profile than the PSF to model the stars to better account for source crowding and confusion.

The observed, model and residual images of all bands are displayed in Figs.~\ref{galfit1.fig} and \ref{galfit2.fig}. The residuals are all on very small levels compared with the significant brightness of the star cluster. We also have calculated the total residual fluxes within 5 pixels from the cluster centre, which are then added to \textit{Cluster~A}'s model fluxes in the corresponding bands. The residual fluxes, however, are all smaller than the fitting uncertainties.

\section{Decontamination of \textit{WFPC2}/\textit{F658N} photometry}
\label{line.sec}

Due to the poor spatial resolution, the source in the \textit{WFPC2/F658N} image includes the contribution not only from \textit{Cluster~A} but also from \textit{S1} and \textit{S2} in its close proximity. To study the properties of \textit{Cluster~A}, we must estimate and remove the contribution of \textit{S1} and \textit{S2} properly.

To this end, we first linearly interpolate the continuum fluxes of \textit{S1} and \textit{S2} in the \textit{WFC3/F657N} and \textit{WFPC2/F658N} bands from their \textit{F555W}- and \textit{F814W}-band fluxes [\textit{F555W}(\textit{S1})~=~23.67~$\pm$~0.03~mag; \textit{F814W}(\textit{S1})~=~22.07~$\pm$~0.01~mag; \textit{F555W}(\textit{S2})~=~23.42~$\pm$~0.02~mag; \textit{F814W}(\textit{S2})~=~21.68~$\pm$~0.01~mag]. We then create synthetic spectrum by adding emission lines of H$\alpha$ and [N~{\small II}]~$\lambda\lambda$6548, 6584 to the continuum; the line wavelengths have been Doppler-shifted with a recession velocity of 990~km~s$^{-1}$ \citepalias{M15} and we assume the flux ratio of ([N~{\small II}]~$\lambda$6548)/([N~{\small II}]~$\lambda$6584) to be 1/3 \citep{agn2.ref}. We adjust the fluxes of H$\alpha$ and [N~{\small II}] lines until the synthetic magnitudes of \textit{S1} and \textit{S2} in the \textit{WFC3/F657N} band match those derived from GALFIT within errors [\textit{F657N}(\textit{S1})~=~21.61~$\pm$~0.06~mag; \textit{F657N}(\textit{S2})~=~20.85~$\pm$~0.04~mag]. The brightness of \textit{S1} and \textit{S2} in the \textit{WFPC2/F658N} band can then be estimated with their synthetic magnitudes.

We find that the results are weakly dependent on the flux ratio of ([N~{\small II}]~$\lambda$6584)/(H$\alpha$~$\lambda$6563). By changing this ratio from 0.01 to 1.0, the \textit{WFPC2/F658N}-band magnitude varies over a range of 19.70--19.83~mag for \textit{S1} and 20.74--20.874 for \textit{S2}. As a result, \textit{Cluster~A}'s contribution lies within 19.47--19.62~mag. We take the median value (19.54~mag) as the decontaminated magnitude of \textit{Cluster~A} and the half range (0.07~mag) as the uncertainty caused by the unknown ([N~{\small II}]~$\lambda$6584)/(H$\alpha$~$\lambda$6563) flux ratio. This uncertainty is combined with the model fitting error as the total uncertainty. Thus, we obtain the decontaminated magnitude of \textit{Cluster~A} as \textit{F658N}~=~19.54~$\pm$~0.09~mag.

%To this end, we first estimate the continuum 

\bsp	% typesetting comment
\label{lastpage}
\end{document}